\documentclass[english,aps,pra,twocolumn,showpacs]{revtex4-1}
\usepackage[T1]{fontenc}
\usepackage[latin9]{inputenc}
\usepackage{amsmath}
\usepackage{graphicx}

\makeatletter

\providecommand{\tabularnewline}{\\}

\makeatother

\usepackage{babel}
\begin{document}

\title{Supercell convergence of charge-transfer energies in pentacene molecular crystals from constrained DFT}

\author{David H. P. Turban}

\affiliation{Cavendish Laboratory, 19 J.J. Thomson Avenue, Cambridge, CB3 0HE, UK.}

\author{Gilberto Teobaldi}
\affiliation{ Stephenson Institute for Renewable Energy and Department of Chemistry,
The University of Liverpool, Liverpool, L69 7ZF, UK.}

\author{David D. O'Regan}

\affiliation{CRANN, AMBER, and School of Physics, 
Trinity College Dublin, Dublin 2, Ireland.}

\author{Nicholas D. M. Hine}

\affiliation{Department of Physics, University of Warwick, Gibbet Hill Road,
Coventry, CV4 7AL, UK.}

\date{\today}
\begin{abstract}
Singlet fission (SF) is a multi-exciton generation process that could
be harnessed to improve the efficiency of photovoltaic devices. 
Experimentally, systems derived from the pentacene molecule have 
been shown to exhibit ultrafast SF with high yields. 
Charge-transfer (CT) configurations are
likely to play an important role as intermediates in the SF process in
these systems. In molecular crystals, electrostatic screening effects
and band formation can be significant in lowering the energy of CT
states, enhancing their potential to effectively participate in SF.
In order to simulate these, it desirable to adopt a computational 
approach which is acceptably accurate, relatively inexpensive,
which and scales well to larger systems, 
thus enabling the study of screening effects. 
We propose a novel, electrostatically-corrected 
constrained Density Functional Theory (cDFT) approach  
as a low-cost solution to the calculation of CT 
energies in molecular crystals such as pentacene.
Here we consider an implementation in the context of the
ONETEP linear-scaling DFT code, but our electrostatic correction
method is in principle applicable in combination with any constrained
DFT implementation, also outside the linear-scaling framework.
Our newly developed method allows 
us to estimate CT energies in the infinite crystal limit, 
and with these to validate the accuracy of the cluster approximation.
\end{abstract}
\maketitle

\section{Introduction}

\begin{figure}
\includegraphics[width=1\columnwidth]{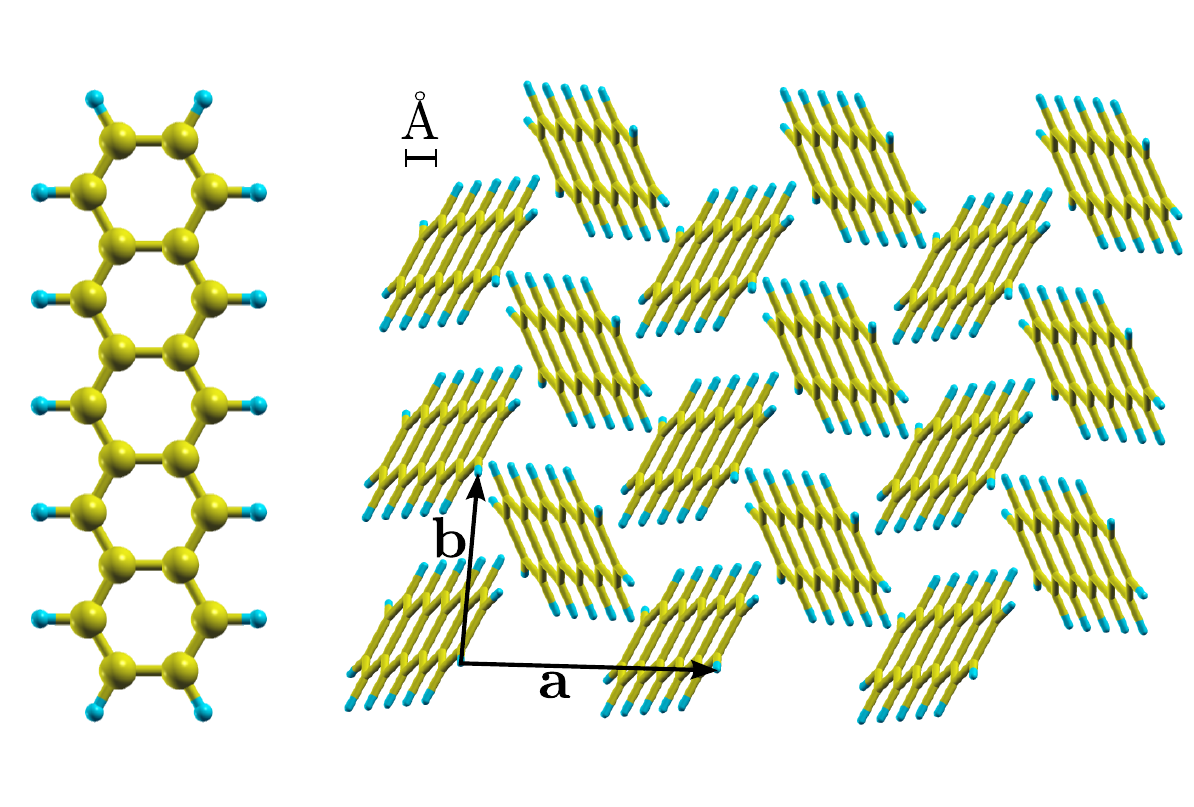}

\caption{The pentacene ($\text{C}_{22}\text{H}_{14}$) single molecule and molecular
crystal (S-phase \citep{sharifzadeh_quasiparticle_2012}). The unit
cell contains two molecules. The third lattice vector $\mathbf{c}$
points out of the page.\label{fig:pentacene-lattice}}
\end{figure}

Singlet fission (SF) is a multiple-exciton-generation process that
is of great interest for potential applications in photovoltaics \citep{smith_singlet_2010,smith_recent_2013,lee_singlet_2013}.
Crystalline pentacene is a material that exhibits highly efficient SF on an ultrafast timescale of around 80~fs \citep{wilson_ultrafast_2011}. 
The system has been investigated in a number of experimental 
\citep{rao_exciton_2010,chan_observing_2011,marciniak_ultrafast_2007,marciniak_ultrafast_2009}, theoretical 
\citep{beljonne_charge-transfer_2013,berkelbach_microscopic_2013,zimmerman_mechanism_2011,zimmerman_singlet_2010,zeng_low-lying_2014}, and combined 
\citep{congreve_external_2013,bakulin_real-time_2016} studies.
Pentacene is a five-membered linear polyacene and forms molecular
crystals with a characteristic `herringbone' lattice (Fig.~\ref{fig:pentacene-lattice}).
Intermolecular charge-transfer (CT) states spanning nearest-neighbour
molecules are thought to play a central role in the ultrafast SF process
in pentacene and similar molecules \citep{berkelbach_microscopic_2013-1,berkelbach_microscopic_2013,berkelbach_microscopic_2014,chan_quantum_2013}.
In molecular crystals, neighbouring pairs of molecules (dimers) undergoing
charge-transfer are embedded in a complex molecular environment where
screening and hybridisation effects are important.

Angular-momentum preserving CT states are difficult to 
access with experimental techniques as
they are optically dark, making it challenging to measure accurate excitation energies
\citep{sebastian_charge_1981,sebastian_charge-transfer_1983}. They
also pose a challenge to electronic structure theory. For example,
linear-response time-dependent DFT with local exchange is known to
perform poorly on CT-like states, severely underestimating their energies
\citep{dreuw_long-range_2003}. This stems from the fact that local
functionals are unable to describe the long-ranged electron-hole interaction
correctly, a problem that is related to, and which compounds, the well-known band-gap error
of the underlying ground-state DFT calculation.
A possible remedy is the use of long-range corrected functionals
with asymptotically correct exchange \citep{tawada_long-range-corrected_2004}.
However, this comes at the expense of introducing additional parameters
and of a greatly increased computational cost. It is thus desirable and timely to 
construct a low-cost method that scales well to large system sizes and 
complex environments, and which simultaneously describes the electrostatic
and quantum mechanical features of CT states with reasonable accuracy.

In this work, we make use of constrained DFT (cDFT)
\citep{dederichs_ground_1984,wu_direct_2005,wu_constrained_2006,kaduk_constrained_2012}
in combination with linear-scaling DFT (as implemented in the ONETEP
code \citep{skylaris_introducing_2005}), applying it to intermolecular
charge-transfer in two nearest-neighbour dimers taken from the pentacene
crystal structure. The cDFT method has been applied to a wide variety of
molecular systems, to date, in the context of CT excitation energies
\citep{rezac_robust_2015,vaissier_influence_2015,zheng_ab_2012,zheng_solvated_2013}, 
electronic couplings \citep{kubas_electronic_2014,kubas_electronic_2015,si_theoretical_2012},
electron transfer
\citep{aikawa_theoretical_2015,eisenmayer_proton_2013,yu_electrochemical_2012,siefermann_atomic-scale_2014} and molecular dynamics 
\citep{rezac_robust_2012,oberhofer_charge_2009}.
A largely unresolved issue in this context, however,
 is that of achieving supercell convergence of CT
excitations in extended models suitable for capturing the screening
and hybridisation effects encountered in realistic systems.
A solution to this problem, such as that which we presently propose,
is then readily transferable to a range of 
complex systems of technological interest, not only in the context of
photovoltaics, but also organic electronics
\citep{sun_introduction_2008,forrest_path_2004,zhao_25th_2013} 
and spintronics 
\citep{naber_organic_2007}.

We first calculate CT energies for the dimers in isolation, and we subsequently
include screening effects by embedding such dimers in a small cluster
of neighbours, and in supercells of the crystal. Supercell calculations
allow us to approach the infinite limit using a correction scheme
that eliminates the spurious dipole-dipole interactions between periodic
images of the simulation cell. The only inputs required for this correction
are the intrinsic dipole of the CT configuration and the dielectric
tensor of the crystal. The latter is obtained from a density functional
perturbation theory (DFPT) calculation \citep{baroni_phonons_2001}.
We find that a single parameter, fit to the results of a series of
calculations on different supercells,
is sufficient to correct for the overestimation of electrostatic
screening as a result of the aforementioned band-gap problem of DFT. 
The isolated calculations facilitate a comparison of the cDFT method
with higher-level theory results from the literature \citep{coto_low-lying_2014}. 
In addition
comparison between clusters and the infinite limit enables us to directly
confirm the validity of the cluster approximation.

\section{Methods}

\subsection{The ground state: linear-scaling DFT}

In order to carry out ground and excited state calculations on large clusters
and supercells, we use linear-scaling DFT as implemented in the
ONETEP code \citep{skylaris_introducing_2005}. This LS-DFT
methodology is based on the single-electron density matrix
$\rho(\mathbf{r},\mathbf{r}')$ rather than Kohn-Sham orbitals
$\psi_{i}(\mathbf{r})$. The density
matrix is expanded in a basis of localised, atom-centred functions
$\phi_{\alpha}(\mathbf{r})$ called NGWFs (non-orthogonal generalised
Wannier functions) \citep{skylaris_nonorthogonal_2002}:
\begin{equation}
\rho(\mathbf{r},\mathbf{r}')=\sum_{i}\psi_{i}(\mathbf{r})f_{i}\psi_{i}(\mathbf{r}')=\sum_{\alpha\beta}\phi_{\alpha}(\mathbf{r})K^{\alpha\beta}\phi_{\beta}(\mathbf{r}'),
\end{equation}
where $K^{\alpha\beta}$ is called the density kernel. The NGWFs are
strictly truncated at a chosen localisation radius, which is a 
convergence parameter. The computational
effort of traditional DFT methods, based on 
manipulation of Kohn-Sham eigenstates,
 inevitably scales as $\mathcal{O}(N^{3})$, where
$N$ denotes the number of electrons. This is because
there are $\mathcal{O}(N)$ eigenstates represented via
$\mathcal{O}(N)$ basis functions, which have to
be kept mutually orthogonal to $\mathcal{O}(N)$ other eigenstates.
By contrast, in a density matrix representation, it is possible
to achieve overall $\mathcal{O}(N)$ scaling if the density kernel
is truncated at some cutoff radius such that it is a sparse matrix.
This exploits the `near-sightedness' of electronic structure in systems
with a gap \citep{kohn_nearsightedness_2008}. Instead of imposing
orthogonality explicitly on Kohn-Sham states, it is necessary to constrain
the density matrix to be idempotent and have a trace equal to the
number of electrons $N$. In ONETEP, a nested loop optimisation
scheme, utilising a conjugate gradients
algorithm, is used to minimise the total energy with
respect to both $K^{\alpha\beta}$ (subject to the constraints of
idempotency and normalisation), and the set of NGWFs $\{\phi_{\alpha}(\mathbf{r})\}$.
This approach has been shown to provide total energies and
forces in $\mathcal{O}(N)$ effort with systematically controllable
accuracy equivalent to that of a plane-wave basis \citep{skylaris_achieving_2007}.
This is possible despite using a minimal number of NGWFs (i.e., 
typically one per hydrogen atom, four per carbon atom),
since the NGWFs are optimised in an
underlying variational basis set of `psinc' functions (delta functions with
limited spectral range), which are fully equivalent to plane-waves.

\subsection{Constrained DFT\label{sub:Constrained-DFT}}

\begin{figure}
\includegraphics[width=1\columnwidth]{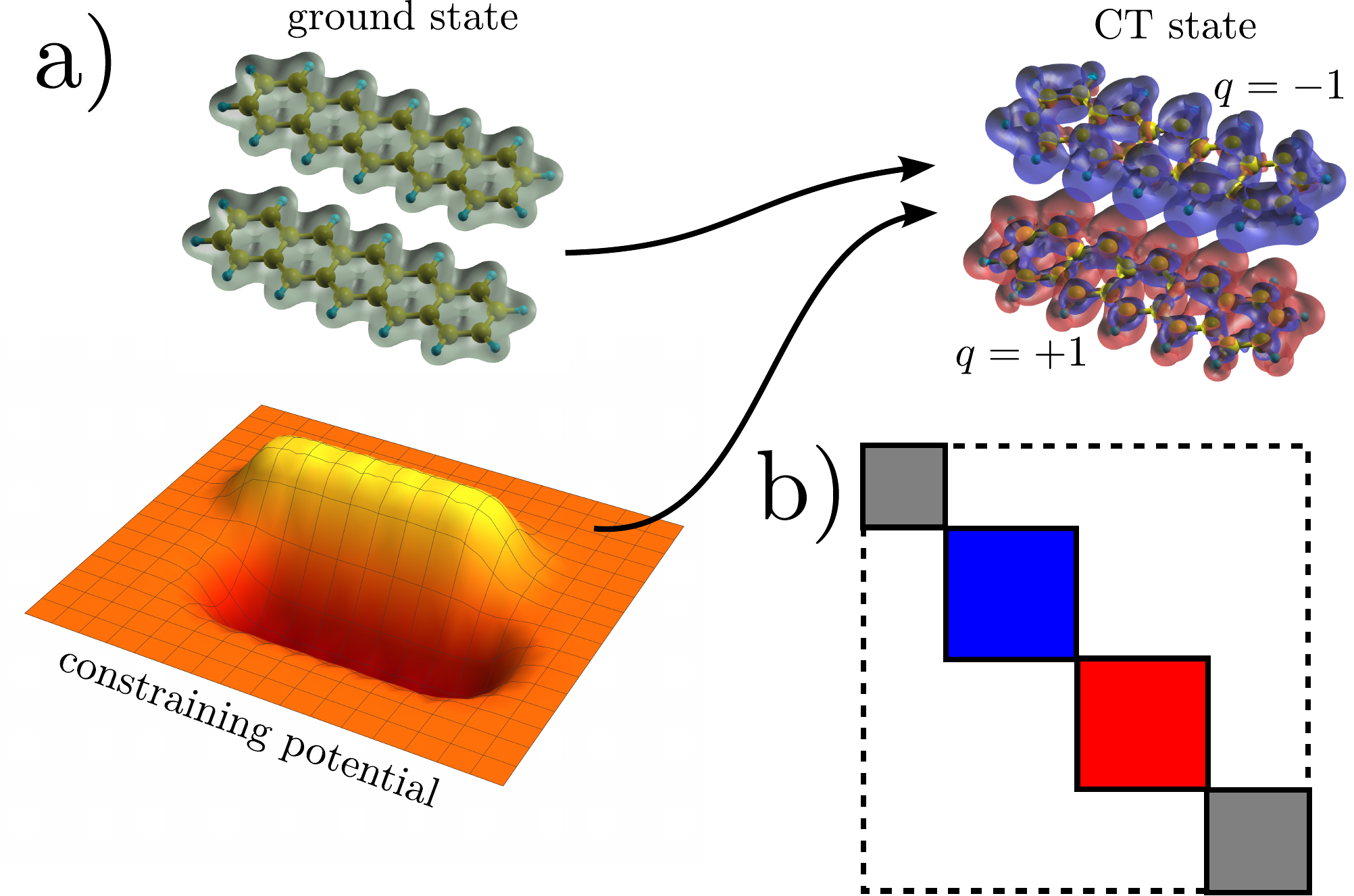}

\caption{a) Schematic of the cDFT scheme used in this work: A
nonlocal constraining potential (illustrated by 2D potential energy surface)
 constructed from atom-centred
functions is applied to the single-electron density matrix.
This causes charge to redistribute to obey chosen population
constraints, and allows the description of CT excitations
within the framework of standard DFT. b) Block scheme of truncated
NGWF overlap matrix to ensure site-localisation of contravariant
duals. Blue and red denote the constrained sites, gray the remaining
system.\label{fig:cDFT-in-ONETEP:}}

\end{figure}

In constrained DFT (cDFT) \citep{dederichs_ground_1984,wu_direct_2005,wu_constrained_2006}
the DFT total energy functional is augmented with terms that impose
desired constraints on the charge (and/or spin) density of a system.
While these constraints can take several forms, in this work we impose
them using monomer-localised projection operators to partition 
the density.
This gives a total functional of the form:
\begin{equation}
W=E_{\text{DFT}}+\sum_{\text{sites }I}V_{I}\left(\text{Tr}[\hat{P}_{I}\hat{\rho}]-N_{I}\right).\label{eq:cDFT-func}
\end{equation}
Here, the $V_{I}$ are Lagrange multipliers that enforce occupancy
targets $N_{I}$ on specific sites in the system, which
are defined via projectors $\hat{P}_{I}$.
The sites in question may, generally, be atoms, groups of atoms or entire molecules.
For example, if one aims to describe an intermolecular CT state, each
of the two molecules involved constitutes a site. The Lagrange multipliers
$V_{I}$ act as artificial constraining potentials that cause charge
to move around the system (cf. Fig.~\ref{fig:cDFT-in-ONETEP:}a).
These potentials are optimised in-situ, via a further conjugate
gradients algorithm nested between kernel optimisation and NGWF
optimisation, and iterated until the population targets
$N_{I}$ for the chosen sites are met. In the case of intermolecular
CT states, these targets are, respectively,
one fewer charge on the donor molecule and
one additional charge on the acceptor, relative to the ground state.

Within the LS-DFT framework it is a natural choice to employ the 
aforementioned localised NGWFs to define site projectors \citep{oregan_projector_2010}.
In this work we employ a fixed set of NGWFs from a ground-state calculation for this purpose:
\begin{equation}
\hat{P}_{I}=\sum_{\alpha\in I}\left|\phi^{\alpha}\right\rangle \left\langle \phi_{\alpha}\right|,
\end{equation}
where the sum $\alpha \in I$ refers to NGWFs centered on atoms
belonging to site $I$. Here, subscript indices are used to
describe the standard covariant functions $|\phi_\alpha\rangle$,
while superscript indices refer to their contravariant
duals $|\phi^\alpha\rangle$, which obey $\langle \phi_\alpha |
\phi^\beta \rangle = \delta_{\alpha\beta}$. See 
Ref.~\citep{artacho_nonorthogonal_1991, oregan_linear-scaling_2012} 
for further discussion on this topic.
A complication in the definition of site projection operators arises from the
fact that the NGWFs are not orthonormal (this is true even of atomic orbitals, in the 
case of sites comprising more than one atom),
meaning that the duals $\phi^{\alpha}$ are not the same as the NGWFs.
Instead, they are defined via the inverse
of the NGWF overlap matrix $S_{\alpha\beta}=\langle\phi_{\alpha}|\phi_{\beta}\rangle$:
\begin{equation}
\left|\phi^{\alpha}\right\rangle =\sum_{\beta} \left|\phi_{\beta}\right\rangle 
\left(S^{-1}\right)^{\beta \alpha}.
\end{equation}
Given that the overlap and inverse overlap matrices can both
be made sparse by appropriately-chosen truncation, it is possible
to construct the inverse in linear-scaling computational
effort using a sparse matrix implementation of Hotelling's
algorithm~\cite{ozaki_efficient_2001}.
Duals constructed using the NGWF overlap matrix for the complete system
are delocalised over that system, 
and  thus present a highly undesirable choice for use in cDFT since
this implies that constraining potentials $V_{I}$ act non-locally
on the charge density, with donor and acceptor subspaces overlapping.
Appropriate localisation of the duals, to the region
of the system of interest for defining a site, is achieved by
by suitably truncating the NGWF overlap matrix before its
inversion, and then defining subspace duals for the purposes of building the
site projection operators via the resulting subspace 
inverse overlap matrix $O_{\alpha\beta}$ instead of the
full $S_{\alpha\beta}$, as described in \citep{oregan_subspace_2011}.
Specifically, a `site-block' scheme is imposed on the 
sparsity of the NGWF overlap matrix before it is inverted.
Here, a block is defined by all NGWFs
associated with a given site. Overlap matrix
elements between NGWFs associated with different sites are set to zero
(cf. Fig.~\ref{fig:cDFT-in-ONETEP:}b). Once this matrix has been
inverted, it retains the same block pattern of sparsity, meaning
that subspace duals are defined as a linear combination of only
those NGWFs on the same constraint site.

When the sites are defined in self-contained manner, thereby,
bi-orthogonality is unavoidably lost between NGWFs and duals localised to different sites,
in the event that these sites overlap to some degree.
This carries the disadvantage the sum of charges 
over a set of such sites,  covering the system,
may not equal the true total charge.
For well-separated donor and acceptor regions such as in the
system at hand, any overestimation of site charge due to the latter effect
is insubstantial in comparison to the dramatic overestimation
incurred by using delocalised duals.
On the other hand, even when the donor and acceptor regions do overlap substantially, unlike methods employing fully delocalized duals our approach ensures that the constraining potentials remain fully localised to their respective regions, with a smooth, non-oscillatory transition at the boundary.

A different approach to cDFT in the context of linear-scaling has been described in Refs.~\citep{ratcliff_toward_2015,ratcliff_fragment_2015}.

In order to obtain energies of CT excitations, we first perform
a ground state DFT calculation. This yields both a total energy for
the ground state and a set of converged ground-state NGWFs which
are subsequently used as cDFT projectors. To define the population
targets for the cDFT run we simply add $\pm1$ to the ground state
populations of the appropriate sites (acceptor: $+1$; donor: $-1$). The 
difference between the constrained total energy and the ground state energy 
yields the (vertical) CT excitation energy. Since we are interested in ultrafast
processes like singlet fission where nuclear relaxation in the excited state is less
significant, we restrict our attention to vertical excitation energies. 
In general, a geometry optimisation in the excited state would be required in order to correctly describe longer-lived CT states.

\subsection{Computational Details}

For all calculations we employ the LDA functional and norm-conserving
pseudopotentials. The energy cutoff is chosen as 750~eV. We use 1 NGWF 
per hydrogen atom and 4 NGWFs per carbon atom. For the localisation radius 
of the NGWFs a value of 10 Bohr is chosen.
Using these parameters the total energy is converged to 1~meV/atom at 10~Bohr NGWF radius
compared to 14~Bohr, and to around 25~meV/atom at 750~eV cutoff compared to 1250~eV.
All NGWFs are initialised to pseudoatomic
orbitals \citep{ruiz-serrano_pulay_2012} and then optimised in-situ
in terms of the underlying psinc basis. The density kernel $K^{\alpha\beta}$
is not truncated in this work as all systems are small enough that
sparse matrix algebra is only a minor component of the total computational
effort. For the later Density Functional Perturbation Theory calculations,
we utilise the CASTEP plane-wave DFT code \citep{clark_first_2009} with the
same pseudopotentials and cutoff energy.
The DFPT calculations are performed with 12 k-points, corresponding
to a maximum k-point spacing of 0.05~1/\AA.

The S-phase molecular crystal structure considered here has two
molecules per primitive cell and triclinic ($P-1$) space group
symmetry. The lattice parameters are given by $a=7.90$~\AA, $b=6.06$~\AA,
$c=16.01$~\AA, and $\alpha=101.9^{\circ}$, $\beta=112.6^{\circ}$,
$\gamma=85.8^{\circ}$ \citep{campbell_crystal_1962}.
Optimised molecular geometries are
taken from Ref.~\citep{sharifzadeh_quasiparticle_2012} in order
to facilitate comparison of our calculations with Ref.~\citep{coto_low-lying_2014}
where high-level CASPT2/CASSCF and GW/BSE calculations were performed
using the same geometries.

For calculations on isolated dimers and clusters we employ open boundary
conditions. This is achieved by putting the dimers in a large simulation
box and truncating the Coulomb interaction at large distances to eliminate
electrostatic interactions between periodic images \citep{hine_electrostatic_2011}.
The calculations on supercells of the pentacene crystal use periodic
boundary conditions.

\section{Dimer \& Cluster Calculations}

\begin{figure}
\includegraphics[width=1\columnwidth]{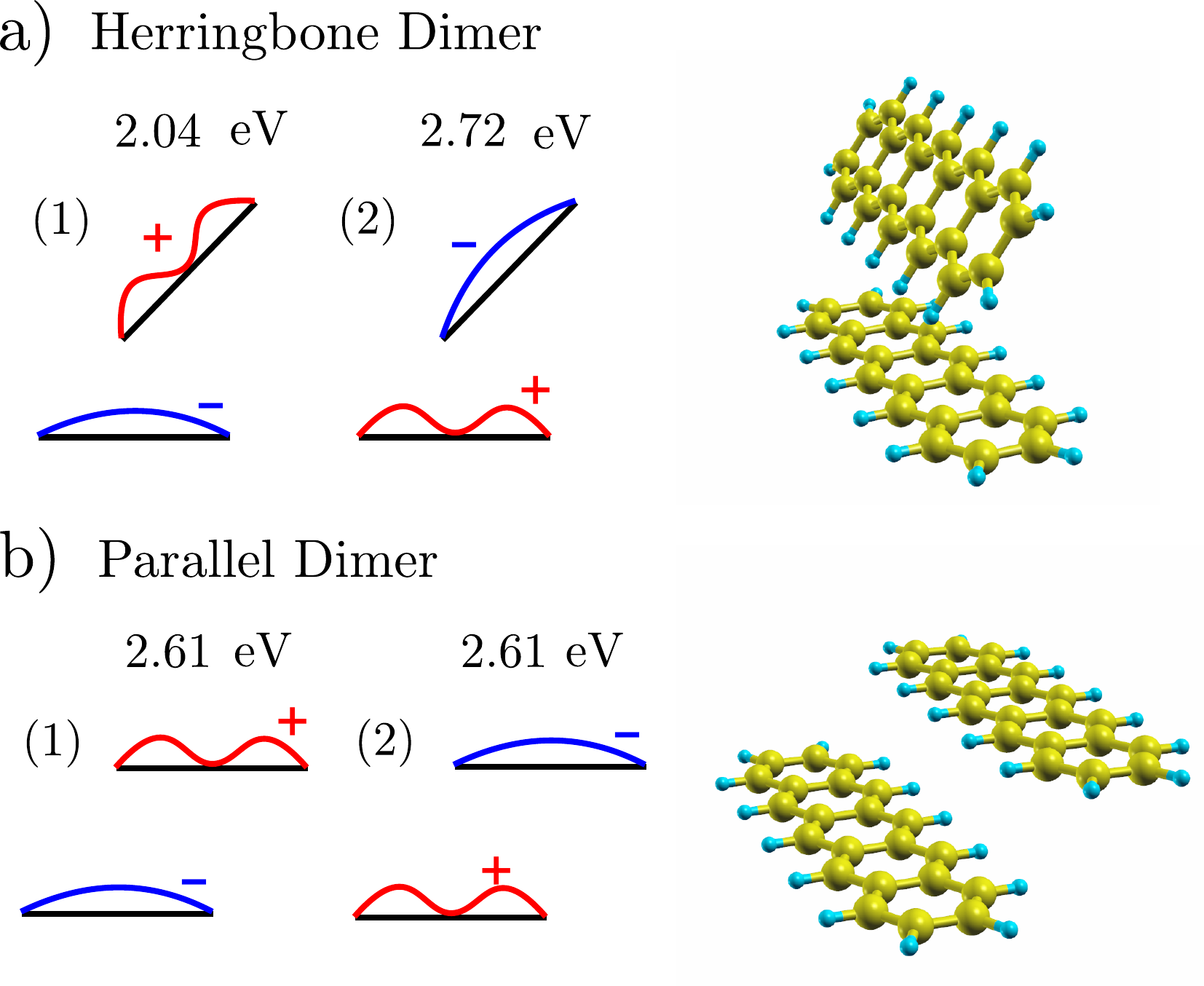}

\caption{Dimer geometries and CT excitation energies from cDFT (quoted in eV).
The significant energy gap between the two CT states in the herringbone
dimer can be rationalised by considering the different charge distributions
of electron and hole, and the geometry. The hole orbital corresponds
to the pentacene HOMO which has a node on the long axis of the molecule.
The electron orbital (LUMO), on the other hand, does not feature such
a node. The partial alignment of the upper molecule with the dipole
vector means that the bimodal charge distribution on the upper molecule
has a lower Coulomb energy in configuration 1 as compared to configuration
2. In the parallel case the two CT states are related by inversion
symmetry and the energies are degenerate.\label{fig:Dimer-geometries}}
\end{figure}

The molecular geometries of the `herringbone' dimer and the `parallel'
dimer are shown in Fig.~\ref{fig:Dimer-geometries}. The herringbone
dimer represents the unit cell of the pentacene crystal. While the
long axes of the molecules are mostly aligned, there is a rotational
offset around the same long axis between the units. In particular,
this means that the two units in the herringbone dimer are not related
by symmetry.

In the parallel dimer, on the other hand, the pentacene molecules
belong to the same sublattice of the crystal and are related by a
translation along lattice vector $\mathbf{b}$ (cf. Fig.~\ref{fig:pentacene-lattice}).
As a result the molecular planes of the molecules are parallel. The
translational correspondence together with the inversion symmetry
of single pentacene molecules mean that the parallel dimer has an
inversion centre, i.e. the units are symmetry-equivalent.

\begin{table}
\begin{tabular}{c||c|c|c}
Configuration & our method & CASPT2/CASSCF & GW/BSE\tabularnewline
\hline 
Herringbone 1 & 2.04 & 2.22 \citep{coto_low-lying_2014} & 1.92 \citep{coto_low-lying_2014}\tabularnewline
Herringbone 2 & 2.72 & 2.55 \citep{coto_low-lying_2014} & 2.60 \citep{coto_low-lying_2014}\tabularnewline
Parallel      & 2.61 & 3.03 \citep{coto_low-lying_2014} & 2.45* \citep{coto_low-lying_2014}\tabularnewline
\end{tabular}

\caption{CT energies (eV) of isolated dimers, comparing our results with higher-level theory.
The authors of Ref.~\citep{coto_low-lying_2014} identify the excitation marked by an asterisk
as a third locally excited state dominated by transitions between the frontier orbitals of 
the monomers. 
However, in a dimer there can only be two states of this kind. Hence, we concluded that
the excitation does in fact have CT character.\label{tab:higher-level-comparison}}
\end{table}

First, we obtain CT energies for the dimers in isolation. The results
are summarised in Fig.~\ref{fig:Dimer-geometries}. The most striking
aspect is the significant energy gap between the two CT configurations
in the herringbone dimer due to their symmetry-inequivalence, as elaborated
in the figure. In the parallel configuration the energies are degenerate
due to inversion symmetry. The excitation energies for the herringbone
configuration are within $\sim$0.2 eV of literature values obtained
with higher-level methods (cf. Table~\ref{tab:higher-level-comparison}). 
For the parallel configuration the discrepancy may be as large as $\sim$0.4 eV,
depending on the method compared to.

We next perform cluster calculations where we surround the dimers
with a small cluster of neighbouring molecules fixed in the geometry
of the molecular crystal (Table~\ref{tab:energy-comparison}). The
4-molecule clusters (144 atoms) only include the CT pair and the two
shared nearest-neighbour molecules in the $\mathbf{a}$-$\mathbf{b}$-plane.
In the 10-molecule clusters (360 atoms), all nearest-neighbour molecules
in the $\mathbf{a}$-$\mathbf{b}$-plane are included. The results
are summarised in Table~\ref{tab:energy-comparison}. We observe
a significant down shift of the mean energy and closing of the relative
gaps as the size of the cluster increases. This is driven by a reduction
of the gap due to the hybridisation and increased electrostatic screening
by neighbouring molecules. It should be noted that the degeneracy
of the two parallel dimer CT states is very slightly lifted in the
clusters (which does not exhibit exact inversion symmetry), but only
within the quoted accuracy. Therefore, we only give a single value
for the CT energy.

From the presented set of calculations alone it is difficult to determine
whether sufficient convergence to the infinite limit has been reached
with the 10-molecule cluster.

\begin{table}
\begin{tabular}{c||c|c|c}
Configuration & isolated (2 mol) & 4-mol cluster & 10-mol cluster\tabularnewline
\hline 
Herringbone 1 & 2.04 (0.80) & 2.16 (0.73) & 2.00 (0.68)\tabularnewline
Herringbone 2 & 2.72 (0.80) & 2.25 (0.73) & 2.04 (0.68)\tabularnewline
Parallel      & 2.61 (1.13) & 2.35 (0.92) & 2.10 (0.87)\tabularnewline
\end{tabular}

\caption{Comparison of CT states for isolated dimers and clusters. The table
quotes the excitation energy of the CT state and the HOMO-LUMO gap
of the ground state configuration in brackets (in units of eV).\label{tab:energy-comparison}}
\end{table}

\section{Supercell Calculations\label{sub:Bulk-Molecular-Crystal:}}

We now consider the dimers embedded not in vacuum, but in the natural
environment of the molecular crystal. This immediately raises the
issue of treating a non-periodic, infinite system in DFT. In practice,
one has to use supercell calculations with periodic boundary conditions.
However, these suffer from finite-size errors which are particularly
pronounced in the case at hand, as can be seen from the large scatter
of 0.1-0.3~eV of the uncorrected CT energies in Fig.~\ref{fig:corr-her1}
(blue bars). This is a consequence of the large dipole moments of
the CT configurations, resulting in significant dipole-dipole interactions
between periodic images. The problem is expected to be even more pronounced
in systems with either larger CT-dipoles, like biological photo-reaction centres,
or smaller polarisabilities (e.g. due to smaller pi-systems).

We address this problem by deriving an energy correction that cancels
the spurious interactions. We apply this correction to a range of
calculations using supercells of varying shapes and sizes to demonstrate
consistency of the method. The largest supercell considered has dimensions
$3\times3\times2$, or 18 unit cells (1368 atoms).

\subsection*{Dipole-dipole correction}

Periodic DFT codes employ the Ewald formula \citep{ewald_berechnung_1921}
to evaluate the electrostatic energy. The central idea is a splitting
of the solution of Poisson's equation $\triangle\varphi=-4\pi\rho$
into parts that converge rapidly in real and reciprocal space, respectively.
The splitting is controlled by an inverse length scale $\eta$, called
Ewald's parameter. 

If metallic boundary conditions are used (i.e. vanishing surface term),
as is commonly the case, the total energy of an isolated system with
a net dipole converges relatively slowly ($\sim V_{\text{cell}}^{-1}$)
with the size of the simulation cell due to dipole-dipole interactions
between periodic images. Makov \& Payne showed that a better degree
of convergence ($\sim V_{\text{cell}}^{-5/3}$) can be achieved by
subtracting a dipole term from the Ewald energy \citep{makov_periodic_1995}:
\begin{equation}
E_{\text{dip}}=-\frac{2\pi}{3V_{\text{cell}}}\cdot\mathbf{P}^{2},\label{eq:surf-vac}
\end{equation}
where $\mathbf{P}$ is the total dipole moment of the simulation cell.
The Makov \& Payne result is only valid for cubic cells. If the aperiodic
system to be studied is embedded in an isotropic dielectric one can
apply the phenomenological approach by Leslie \& Gillan \citep{leslie_energy_1985}.
Here the dipole-dipole interaction is reduced by the dielectric constant
of the dielectric, i.e.
\begin{equation}
E_{\text{dip}}=-\frac{2\pi}{3V_{\text{cell}}}\cdot\frac{|\mathbf{P}|^{2}}{\epsilon}.
\end{equation}

In order to be useful for CT states in pentacene supercells it is
necessary to generalise these results to arbitrary cell shapes and,
crucially, anisotropic dielectrics \citep{mckenna_crossover_2012,blumberger_constrained_2013,murphy_anisotropic_2013}. To achieve the first step we can
employ the following expression by Kantorovich which is valid for
general periodic cells and can obtained by evaluating the Ewald formula
for a periodic lattice of point dipoles \citep{kantorovich_elimination_1999}:
\begin{equation}
E_{\text{dip}}=-\frac{2\eta^{3}}{3\sqrt{\pi}}\cdot|\mathbf{P}|^{2}+\frac{1}{2}\sum_{\alpha,\beta}P_{\alpha}\psi_{\alpha\beta}P_{\beta},
\end{equation}
where $\eta$ is Ewald's parameter, and
\begin{equation}
\psi_{\alpha\beta}=\frac{4\pi}{V_{\text{cell}}}\sum_{\mathbf{k}\ne\mathbf{0}}\frac{k_{\alpha}k_{\beta}}{|\mathbf{k}|^{2}}e^{-|\mathbf{k}|^{2}/4\eta^{2}}-\eta^{3}\sum_{\mathbf{l}\ne\mathbf{0}}H_{\alpha\beta}(\eta\mathbf{l}),
\end{equation}
where $\mathbf{k}$ denotes reciprocal lattice vectors, and $\mathbf{l}$
denotes direct lattice vectors. Furthermore, we have
\begin{equation}
H_{\alpha\beta}(\mathbf{y})=-\delta_{\alpha\beta}h(|\mathbf{y}|)+\frac{y_{\alpha}y_{\beta}}{|\mathbf{y}|^{2}}\left[3h(|\mathbf{y}|)+\frac{4}{\sqrt{\pi}}e^{-|\mathbf{y}|^{2}}\right],
\end{equation}
with
\begin{equation}
h(y)=\frac{2}{\sqrt{\pi}}\cdot\frac{e^{-y^{2}}}{y^{2}}+\frac{\text{erfc}(y)}{y^{3}}.
\end{equation}

If the dipoles are embedded in an anisotropic dielectric we need to
modify Poisson's equation by substituting $\bigtriangleup\rightarrow\boldsymbol{\nabla}^{t}\underline{\underline{\epsilon}}\boldsymbol{\nabla}$,
where $\underline{\underline{\epsilon}}$ is the dielectric tensor
of the medium. Now it turns out that in order to preserve the structure
of the Ewald formula it is necessary to split the charge distribution
anisotropically. This is achieved by inserting the inverse dielectric
tensor in the exponential of the Gaussian smearing function, namely
$\text{exp}(-\eta^{2}|\mathbf{r}|^{2})\rightarrow\text{exp}(-\eta^{2}\mathbf{r}^{t}\underline{\underline{\epsilon}}^{-1}\mathbf{r})/\sqrt{\det\underline{\underline{\epsilon}}}$.
Using this substitution the solution of the (real space) Poisson equation
for the smeared charge can be reduced to the isotropic case by a change
of variables. In the reciprocal space term the denominator transforms
in conjunction with the Poisson equation as $|\mathbf{k}|^{2}\rightarrow\mathbf{k}^{t}\underline{\underline{\epsilon}}\mathbf{k}$.
In addition, the Fourier transform of the modified Gaussian smearing
function is now given by $\text{exp}(-\mathbf{k}^{t}\underline{\underline{\epsilon}}\mathbf{k}/4\eta^{2})$.

All in all, a structurally identical expression for the dipole contribution
is obtained, the only difference being an overall factor of $1/\sqrt{\det\underline{\underline{\epsilon}}}$
and linear transformations of the direct and reciprocal vectors:
\begin{eqnarray}
\mathbf{P} & \rightarrow & \underline{\underline{D}}^{-1}\underline{\underline{C}}^{t}\mathbf{P},\\
\mathbf{l} & \rightarrow & \underline{\underline{D}}^{-1}\underline{\underline{C}}^{t}\mathbf{l},\\
\mathbf{k} & \rightarrow & \underline{\underline{D}}\underline{\underline{C}}^{t}\mathbf{k}.
\end{eqnarray}
Here $\underline{\underline{C}}$ is the (orthogonal) principal axis
transformation that diagonalises $\underline{\underline{\epsilon}}$,
and $\underline{\underline{D}}=\text{diag}(\sqrt{\epsilon_{1}},\sqrt{\epsilon_{2}},\sqrt{\epsilon_{3}})$
with the eigenvalues $\epsilon_{i}$ of $\underline{\underline{\epsilon}}$,
i.e. $\underline{\underline{\epsilon}}=\underline{\underline{C}}\underline{\underline{D}}^{2}\underline{\underline{C}}^{t}$
\citep{fischerauer_comments_1997}. We note that transforming the
lattice vectors in this way necessarily entails a rescaling of the
cell volume, namely $V_{\text{cell}}\rightarrow V_{\text{cell}}/\sqrt{\det\underline{\underline{\epsilon}}}$.

\subsection*{Applying the correction}

\begin{figure}
\includegraphics[width=1\columnwidth]{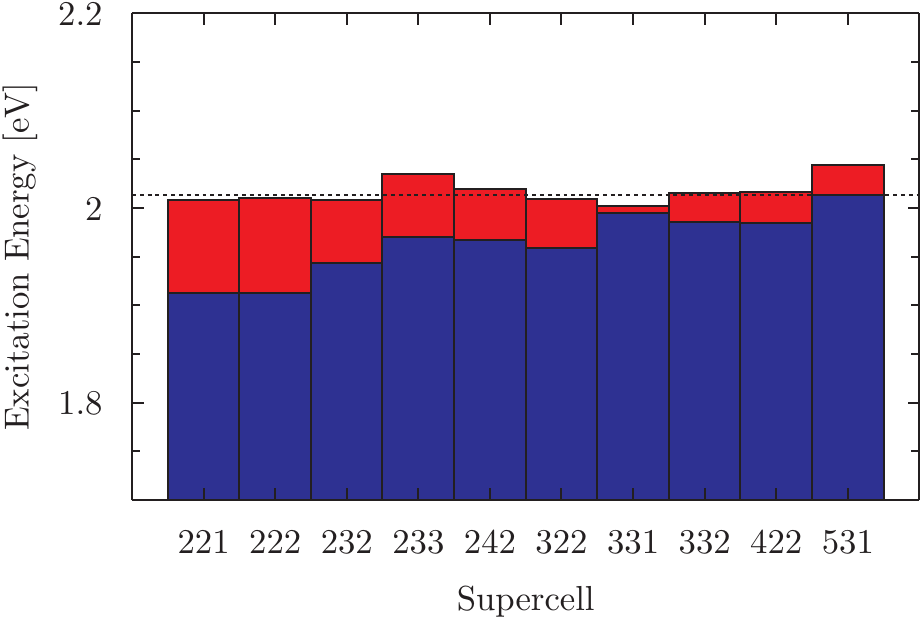} 

\includegraphics[width=1\columnwidth]{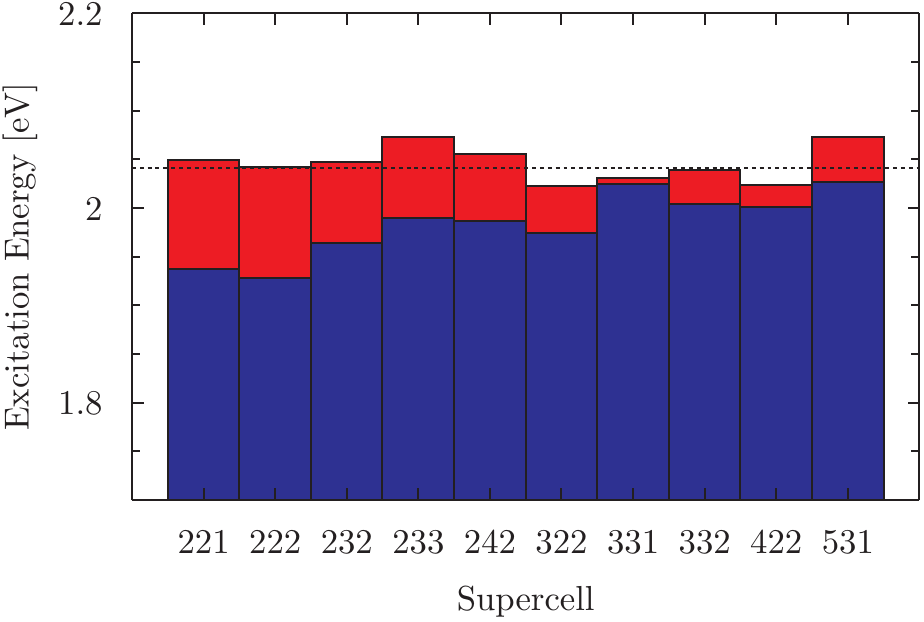}

\includegraphics[width=1\columnwidth]{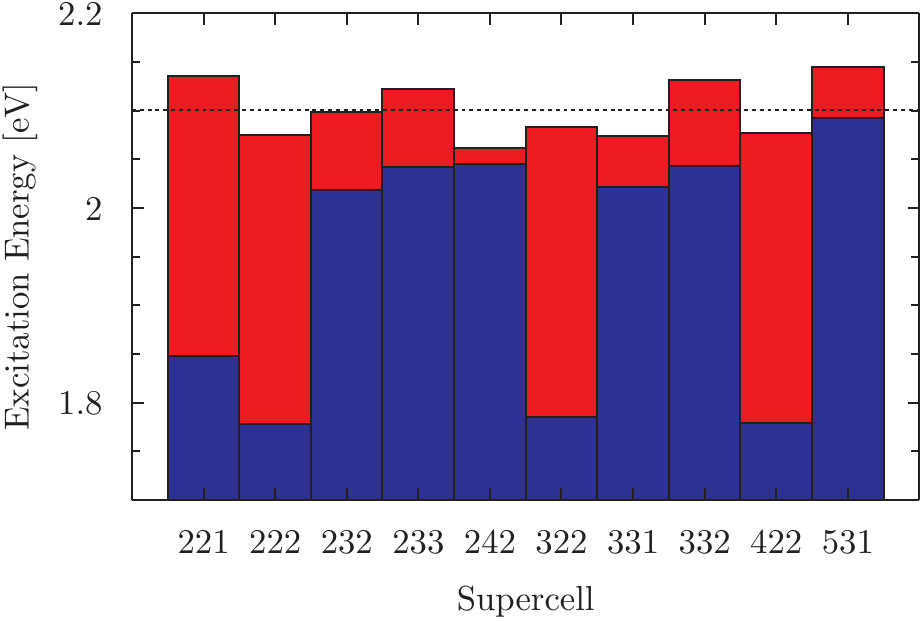}

\caption{\emph{(Top)} Herringbone 1 \emph{(Centre)} Herringbone 2 \emph{(Bottom)}
Parallel. Uncorrected energies (blue) and dipole corrections (red)
for CT states across a range of supercell embeddings. Dashed lines 
indicate corrected mean energies. Note that the
dipole correction is negative for the 531 supercell for both herringbone
configurations.\label{fig:corr-her1}}
\end{figure}
First, we perform a DFPT calculation using the CASTEP code to obtain
the dielectric tensor for the primitive cell of the pentacene crystal:
\[
\underline{\underline{\epsilon}}_{\text{DFPT}}=\left[\begin{array}{ccc}
3.48 & -0.18 & -0.12\\
-0.18 & 3.14 & 0.19\\
-0.12 & 0.19 & 5.61
\end{array}\right].
\]
The actual dielectric tensor is assumed to be obtained by a uniform
scaling $\underline{\underline{\epsilon}}=c\cdot\underline{\underline{\epsilon}}_{\text{DFPT}}$.
The scaling accounts for the overestimation of screening due to the
band-gap error, as previously discussed.
By using a single scaling parameter $c$ we employ the simplifying assumption that
the overscreening of DFPT is isotropic. The dipole moments were taken
from dimer cDFT calculations in vacuum (atomic units):
\begin{align*}
\mathbf{P}_{\text{her1}} & = & (1.45,\,-6.70,\,-2.07)^{t}, & \;\;\left|\mathbf{P}_{\text{her1}}\right| & = &  & 7.16\\
\mathbf{P}_{\text{her2}} & = & (-1.19,\,7.58,\,2.33)^{t}, & \;\;\left|\mathbf{P}_{\text{her2}}\right| & = &  & 8.02\\
\mathbf{P}_{\text{par1/2}} & = & \pm(-7.89,\,4.23,\,1.85)^{t}, & \;\;\left|\mathbf{P}_{\text{par1/2}}\right| & = &  & 9.14
\end{align*}
The dipole correction is applied to the supercell energies as follows:
\begin{equation}
E_{\text{tot}}=E_{\text{cDFT}}-E_{\text{dip}}.\label{eq:etot}
\end{equation}
Minimising the combined standard deviation across all three dimer
configurations yields a best-fit value $c=0.378$. The effect of the
correction using this value of $c$ for the three cases is shown graphically
in Fig.~\ref{fig:corr-her1}. We note that the value of $c$ is in
approximate agreement with the ratio of the DFT gap of $\sim0.8$~eV
and to the quasiparticle gap of $\sim2.4$~eV calculated with
many-body perturbation theory \citep{sharifzadeh_quasiparticle_2012},
as might be expected.

Having applied Eq.~\eqref{eq:etot}, for the herringbone dimer we
obtain a corrected mean energies of 2.01~eV and 2.04~eV, respectively.
The parallel dimer yields a mean energy of 2.10 eV. The spread of
corrected energies is down to 0.03-0.08~eV, a reduction by a factor
of 3-4, demonstrating the success of the method. The significantly
larger spread of values in the parallel dimer compared to the herringbone
dimer is a result of enhanced dipole interaction between periodic
images. This is due to the fact that the dimer (and hence dipole)
is aligned with the crystal lattice (along lattice vector $\mathbf{b}$,
cf. Fig.~\ref{fig:pentacene-lattice}).

\section{Discussion}

As elaborated above, the empirical parameter $c$ in the method accounts
for the overscreening of DFPT with local functionals. There are additional
uncertainties due to the fact that the separation of periodic images
is small, implying that the supercell calculations still exhibit a
relatively high density of electron-hole pairs. This can modify the
dielectric properties. Further errors may stem from higher-order electrostatic
corrections as a result of the fact that the CT configurations considered
take up a significant portions of the supercells and hence do not
constitute perfectly point-like dipoles. All these effects are integrated
in the $c$ parameter. The remaining spread of energies is presumably
related to residuals of these sources of error which cannot be eliminated
with the single $c$ parameter.

It is apparent that the dipole-corrected energies are essentially
degenerate with those obtained for the 10-molecule cluster, within
the quoted precision of 10~meV (cf. Table~\ref{tab:energy-comparison}).
Presuming this holds for still-larger clusters, this result demonstrates
the validity of the cluster approximation for molecular crystals of
pentacene and likely a range of similar molecules (eg. tetracene).
In case where the molecular unit of a molecular crystal has a net
dipole, the cluster approach in vacuum would incur significant difficulties
due to the unscreened net-dipole \citep{lever_electrostatic_2013}.
The current approach, in employing periodic boundary conditions, would
not encounter such problems.

We observe that aggregation has a twofold effect on CT energies in
the pentacene molecular crystal: an overall downshift, and an assimilation
of the different CT configurations considered. The latter is particularly
striking in the herringbone dimer, where the initial splitting of
nearly 0.7~eV in isolation is reduced to only a few 10~meV when
embedded in the crystal. Interestingly, the alignment of the three
CT energies also changes. In the isolated dimers, the parallel CT
energy is situated between the two herringbone energies, whereas in
the crystal the parallel energy is slightly above the two (now nearly
degenerate) herringbone energies.

We also note that although our calculations employ the computationally
cheap LDA functional they do not appear to suffer from a systematic
underestimation of excitation energies relative to those calculated
with higher levels of theory.
This observation can be rationalised by noting that in cDFT the excited electron
orbital is fully occupied. No DFT eigenvalues of unoccupied orbitals which are
subject to large systematic underestimation enter the calculation.
TDDFT with local exchange requires
computationally expensive and parameter-dependent range-separated
hybrid functionals in order to yield CT energies that are not severely
underestimated \citep{dreuw_long-range_2003,tawada_long-range-corrected_2004}.
By contrast, using the novel combination of linear-scaling methodology with 
projector-based cDFT, we are able to perform relatively cheap calculations
which scale well to large system sizes (1368 atoms, in this work).

We further note that the relatively low CT energies in the crystal
of just above 2 eV put them on the lower end of experimental estimates
\citep{sebastian_charge_1981}. They are in line with previous theoretical
results indicating a significant admixture of CT-like components into
the lowest singlet exciton in pentacene \citep{tiago_ab_2003,cudazzo_excitons_2012,sharifzadeh_low-energy_2013}.
The low energies also lend support to the notion that a CT-mediated
`superexchange' mechanism can play a crucial role in ultrafast fission
\citep{berkelbach_microscopic_2013-1,berkelbach_microscopic_2013,berkelbach_microscopic_2014,chan_quantum_2013}.

Our procedure should also be transferable to other systems in which CT states
are situated in a complex screening environment. These include (but are not
limited to) a variety of organic materials for organic photovoltaics and 
optoelectronics/spintronics.

At this point there remain a number of limitations that need to be addressed by 
future work. Our method as presented in this work relies on the system in question
exhibiting sufficiently homogeneous dielectric properties which can be approximated
by a single dielectric tensor. Furthermore, the cDFT approach necessitates prior
knowledge of the excited state structure such that appropriate donor and acceptor
regions can be defined. Another current drawback is the requirement of performing
calculations on a range of supercells for the purpose of the single parameter fit,
increasing computational cost.
This limitation may be overcome by using a more accurate functional (or
higher-level theory than DFT) for the response calculation, provided that the unit
cell is not so large as to make this computationally infeasible.

\section{Conclusions}

In this work we have demonstrated the application of (linear-scaling)
constrained DFT to charge-transfer states in the pentacene molecular
crystal. Our results for isolated dimers are in reasonable agreement
with higher-level theory calculations from the literature. Furthermore,
we have used cluster calculations to illustrate the transition to
the crystal limit, showing that CT energies are lowered both by screening
and the formation of bands. We have devised a scheme based on periodic
supercell calculations in combination with a dipole correction in
order to establish the limit of the infinite molecular crystal. Our
dipole correction method is novel in that it is applicable to very general
systems (non-cubic with anisotropic dielectric properties). 
It is also of significant interest that in
spite of the high-polarity of the CT configurations, the excellent agreement 
between results for the cluster approximation and for the crystal limit reveals 
unexpectedly high screening capability for pentacene, with important 
consequences for the modelling of pentacene interfaces and film stuctural 
imperfections.

\begin{turnpage}

\end{turnpage}

\begin{ruledtabular}
\end{ruledtabular}

\begin{turnpage}

\end{turnpage}

\begin{acknowledgments}
\appendix
DHPT and NDMH would like to gratefully acknowledge financial support
from the Winton Programme for the Physics of Sustainability. In addition,
DHPT gratefully acknowledges financial support from the
Engineering and Physical Sciences Research Council, as well
as access to computing resources provided by the Cambridge High Performance
Computing Cluster \emph{Darwin}. GT and DDO'R gratefully acknowledge 
the Royal Society and Royal Irish Academy for provision of an 
International Exchange grant.
The underlying data of this publication can be accessed via the following persistent URI: 
\url{https://www.repository.cam.ac.uk/handle/1810/254222}

\end{acknowledgments}

\bibliographystyle{apsrev4-1}
\bibliography{MyLibrary}

\begin{thebibliography}{73}%
\makeatletter
\providecommand \@ifxundefined [1]{%
 \@ifx{#1\undefined}
}%
\providecommand \@ifnum [1]{%
 \ifnum #1\expandafter \@firstoftwo
 \else \expandafter \@secondoftwo
 \fi
}%
\providecommand \@ifx [1]{%
 \ifx #1\expandafter \@firstoftwo
 \else \expandafter \@secondoftwo
 \fi
}%
\providecommand \natexlab [1]{#1}%
\providecommand \enquote  [1]{``#1''}%
\providecommand \bibnamefont  [1]{#1}%
\providecommand \bibfnamefont [1]{#1}%
\providecommand \citenamefont [1]{#1}%
\providecommand \href@noop [0]{\@secondoftwo}%
\providecommand \href [0]{\begingroup \@sanitize@url \@href}%
\providecommand \@href[1]{\@@startlink{#1}\@@href}%
\providecommand \@@href[1]{\endgroup#1\@@endlink}%
\providecommand \@sanitize@url [0]{\catcode `\\12\catcode `\$12\catcode
  `\&12\catcode `\#12\catcode `\^12\catcode `\_12\catcode `\%12\relax}%
\providecommand \@@startlink[1]{}%
\providecommand \@@endlink[0]{}%
\providecommand \url  [0]{\begingroup\@sanitize@url \@url }%
\providecommand \@url [1]{\endgroup\@href {#1}{\urlprefix }}%
\providecommand \urlprefix  [0]{URL }%
\providecommand \Eprint [0]{\href }%
\providecommand \doibase [0]{http://dx.doi.org/}%
\providecommand \selectlanguage [0]{\@gobble}%
\providecommand \bibinfo  [0]{\@secondoftwo}%
\providecommand \bibfield  [0]{\@secondoftwo}%
\providecommand \translation [1]{[#1]}%
\providecommand \BibitemOpen [0]{}%
\providecommand \bibitemStop [0]{}%
\providecommand \bibitemNoStop [0]{.\EOS\space}%
\providecommand \EOS [0]{\spacefactor3000\relax}%
\providecommand \BibitemShut  [1]{\csname bibitem#1\endcsname}%
\let\auto@bib@innerbib\@empty
\bibitem [{\citenamefont {Smith}\ and\ \citenamefont
  {Michl}(2010)}]{smith_singlet_2010}%
  \BibitemOpen
  \bibfield  {author} {\bibinfo {author} {\bibfnamefont {M.~B.}\ \bibnamefont
  {Smith}}\ and\ \bibinfo {author} {\bibfnamefont {J.}~\bibnamefont {Michl}},\
  }\href {\doibase 10.1021/cr1002613} {\bibfield  {journal} {\bibinfo
  {journal} {Chem. Rev.}\ }\textbf {\bibinfo {volume} {110}},\ \bibinfo {pages}
  {6891} (\bibinfo {year} {2010})}\BibitemShut {NoStop}%
\bibitem [{\citenamefont {Smith}\ and\ \citenamefont
  {Michl}(2013)}]{smith_recent_2013}%
  \BibitemOpen
  \bibfield  {author} {\bibinfo {author} {\bibfnamefont {M.~B.}\ \bibnamefont
  {Smith}}\ and\ \bibinfo {author} {\bibfnamefont {J.}~\bibnamefont {Michl}},\
  }\href {\doibase 10.1146/annurev-physchem-040412-110130} {\bibfield
  {journal} {\bibinfo  {journal} {Annu. Rev. Phys. Chem.}\ }\textbf {\bibinfo
  {volume} {64}},\ \bibinfo {pages} {361} (\bibinfo {year} {2013})}\BibitemShut
  {NoStop}%
\bibitem [{\citenamefont {Lee}\ \emph {et~al.}(2013)\citenamefont {Lee},
  \citenamefont {Jadhav}, \citenamefont {Reusswig}, \citenamefont {Yost},
  \citenamefont {Thompson}, \citenamefont {Congreve}, \citenamefont {Hontz},
  \citenamefont {Van~Voorhis},\ and\ \citenamefont {Baldo}}]{lee_singlet_2013}%
  \BibitemOpen
  \bibfield  {author} {\bibinfo {author} {\bibfnamefont {J.}~\bibnamefont
  {Lee}}, \bibinfo {author} {\bibfnamefont {P.}~\bibnamefont {Jadhav}},
  \bibinfo {author} {\bibfnamefont {P.~D.}\ \bibnamefont {Reusswig}}, \bibinfo
  {author} {\bibfnamefont {S.~R.}\ \bibnamefont {Yost}}, \bibinfo {author}
  {\bibfnamefont {N.~J.}\ \bibnamefont {Thompson}}, \bibinfo {author}
  {\bibfnamefont {D.~N.}\ \bibnamefont {Congreve}}, \bibinfo {author}
  {\bibfnamefont {E.}~\bibnamefont {Hontz}}, \bibinfo {author} {\bibfnamefont
  {T.}~\bibnamefont {Van~Voorhis}}, \ and\ \bibinfo {author} {\bibfnamefont
  {M.~A.}\ \bibnamefont {Baldo}},\ }\href {\doibase 10.1021/ar300288e}
  {\bibfield  {journal} {\bibinfo  {journal} {Acc. Chem. Res.}\ }\textbf
  {\bibinfo {volume} {46}},\ \bibinfo {pages} {1300} (\bibinfo {year}
  {2013})}\BibitemShut {NoStop}%
\bibitem [{\citenamefont {Wilson}\ \emph {et~al.}(2011)\citenamefont {Wilson},
  \citenamefont {Rao}, \citenamefont {Clark}, \citenamefont {Kumar},
  \citenamefont {Brida}, \citenamefont {Cerullo},\ and\ \citenamefont
  {Friend}}]{wilson_ultrafast_2011}%
  \BibitemOpen
  \bibfield  {author} {\bibinfo {author} {\bibfnamefont {M.~W.~B.}\
  \bibnamefont {Wilson}}, \bibinfo {author} {\bibfnamefont {A.}~\bibnamefont
  {Rao}}, \bibinfo {author} {\bibfnamefont {J.}~\bibnamefont {Clark}}, \bibinfo
  {author} {\bibfnamefont {R.~S.~S.}\ \bibnamefont {Kumar}}, \bibinfo {author}
  {\bibfnamefont {D.}~\bibnamefont {Brida}}, \bibinfo {author} {\bibfnamefont
  {G.}~\bibnamefont {Cerullo}}, \ and\ \bibinfo {author} {\bibfnamefont
  {R.~H.}\ \bibnamefont {Friend}},\ }\href {\doibase 10.1021/ja201688h}
  {\bibfield  {journal} {\bibinfo  {journal} {J. Am. Chem. Soc.}\ }\textbf
  {\bibinfo {volume} {133}},\ \bibinfo {pages} {11830} (\bibinfo {year}
  {2011})}\BibitemShut {NoStop}%
\bibitem [{\citenamefont {Rao}\ \emph {et~al.}(2010)\citenamefont {Rao},
  \citenamefont {Wilson}, \citenamefont {Hodgkiss}, \citenamefont
  {Albert-Seifried}, \citenamefont {B\"assler},\ and\ \citenamefont
  {Friend}}]{rao_exciton_2010}%
  \BibitemOpen
  \bibfield  {author} {\bibinfo {author} {\bibfnamefont {A.}~\bibnamefont
  {Rao}}, \bibinfo {author} {\bibfnamefont {M.~W.~B.}\ \bibnamefont {Wilson}},
  \bibinfo {author} {\bibfnamefont {J.~M.}\ \bibnamefont {Hodgkiss}}, \bibinfo
  {author} {\bibfnamefont {S.}~\bibnamefont {Albert-Seifried}}, \bibinfo
  {author} {\bibfnamefont {H.}~\bibnamefont {B\"assler}}, \ and\ \bibinfo
  {author} {\bibfnamefont {R.~H.}\ \bibnamefont {Friend}},\ }\href {\doibase
  10.1021/ja1042462} {\bibfield  {journal} {\bibinfo  {journal} {J. Am. Chem.
  Soc.}\ }\textbf {\bibinfo {volume} {132}},\ \bibinfo {pages} {12698}
  (\bibinfo {year} {2010})}\BibitemShut {NoStop}%
\bibitem [{\citenamefont {Chan}\ \emph {et~al.}(2011)\citenamefont {Chan},
  \citenamefont {Ligges}, \citenamefont {Jailaubekov}, \citenamefont {Kaake},
  \citenamefont {Miaja-Avila},\ and\ \citenamefont
  {Zhu}}]{chan_observing_2011}%
  \BibitemOpen
  \bibfield  {author} {\bibinfo {author} {\bibfnamefont {W.-L.}\ \bibnamefont
  {Chan}}, \bibinfo {author} {\bibfnamefont {M.}~\bibnamefont {Ligges}},
  \bibinfo {author} {\bibfnamefont {A.}~\bibnamefont {Jailaubekov}}, \bibinfo
  {author} {\bibfnamefont {L.}~\bibnamefont {Kaake}}, \bibinfo {author}
  {\bibfnamefont {L.}~\bibnamefont {Miaja-Avila}}, \ and\ \bibinfo {author}
  {\bibfnamefont {X.-Y.}\ \bibnamefont {Zhu}},\ }\href {\doibase
  10.1126/science.1213986} {\bibfield  {journal} {\bibinfo  {journal}
  {Science}\ }\textbf {\bibinfo {volume} {334}},\ \bibinfo {pages} {1541}
  (\bibinfo {year} {2011})}\BibitemShut {NoStop}%
\bibitem [{\citenamefont {Marciniak}\ \emph {et~al.}(2007)\citenamefont
  {Marciniak}, \citenamefont {Fiebig}, \citenamefont {Huth}, \citenamefont
  {Schiefer}, \citenamefont {Nickel}, \citenamefont {Selmaier},\ and\
  \citenamefont {Lochbrunner}}]{marciniak_ultrafast_2007}%
  \BibitemOpen
  \bibfield  {author} {\bibinfo {author} {\bibfnamefont {H.}~\bibnamefont
  {Marciniak}}, \bibinfo {author} {\bibfnamefont {M.}~\bibnamefont {Fiebig}},
  \bibinfo {author} {\bibfnamefont {M.}~\bibnamefont {Huth}}, \bibinfo {author}
  {\bibfnamefont {S.}~\bibnamefont {Schiefer}}, \bibinfo {author}
  {\bibfnamefont {B.}~\bibnamefont {Nickel}}, \bibinfo {author} {\bibfnamefont
  {F.}~\bibnamefont {Selmaier}}, \ and\ \bibinfo {author} {\bibfnamefont
  {S.}~\bibnamefont {Lochbrunner}},\ }\href {\doibase
  10.1103/PhysRevLett.99.176402} {\bibfield  {journal} {\bibinfo  {journal}
  {Phys. Rev. Lett.}\ }\textbf {\bibinfo {volume} {99}},\ \bibinfo {pages}
  {176402} (\bibinfo {year} {2007})}\BibitemShut {NoStop}%
\bibitem [{\citenamefont {Marciniak}\ \emph {et~al.}(2009)\citenamefont
  {Marciniak}, \citenamefont {Pugliesi}, \citenamefont {Nickel},\ and\
  \citenamefont {Lochbrunner}}]{marciniak_ultrafast_2009}%
  \BibitemOpen
  \bibfield  {author} {\bibinfo {author} {\bibfnamefont {H.}~\bibnamefont
  {Marciniak}}, \bibinfo {author} {\bibfnamefont {I.}~\bibnamefont {Pugliesi}},
  \bibinfo {author} {\bibfnamefont {B.}~\bibnamefont {Nickel}}, \ and\ \bibinfo
  {author} {\bibfnamefont {S.}~\bibnamefont {Lochbrunner}},\ }\href {\doibase
  10.1103/PhysRevB.79.235318} {\bibfield  {journal} {\bibinfo  {journal} {Phys.
  Rev. B}\ }\textbf {\bibinfo {volume} {79}},\ \bibinfo {pages} {235318}
  (\bibinfo {year} {2009})}\BibitemShut {NoStop}%
\bibitem [{\citenamefont {Beljonne}\ \emph {et~al.}(2013)\citenamefont
  {Beljonne}, \citenamefont {Yamagata}, \citenamefont {Br\'edas}, \citenamefont
  {Spano},\ and\ \citenamefont {Olivier}}]{beljonne_charge-transfer_2013}%
  \BibitemOpen
  \bibfield  {author} {\bibinfo {author} {\bibfnamefont {D.}~\bibnamefont
  {Beljonne}}, \bibinfo {author} {\bibfnamefont {H.}~\bibnamefont {Yamagata}},
  \bibinfo {author} {\bibfnamefont {J.~L.}\ \bibnamefont {Br\'edas}}, \bibinfo
  {author} {\bibfnamefont {F.~C.}\ \bibnamefont {Spano}}, \ and\ \bibinfo
  {author} {\bibfnamefont {Y.}~\bibnamefont {Olivier}},\ }\href {\doibase
  10.1103/PhysRevLett.110.226402} {\bibfield  {journal} {\bibinfo  {journal}
  {Phys. Rev. Lett.}\ }\textbf {\bibinfo {volume} {110}},\ \bibinfo {pages}
  {226402} (\bibinfo {year} {2013})}\BibitemShut {NoStop}%
\bibitem [{\citenamefont {Berkelbach}\ \emph
  {et~al.}(2013{\natexlab{a}})\citenamefont {Berkelbach}, \citenamefont
  {Hybertsen},\ and\ \citenamefont {Reichman}}]{berkelbach_microscopic_2013}%
  \BibitemOpen
  \bibfield  {author} {\bibinfo {author} {\bibfnamefont {T.~C.}\ \bibnamefont
  {Berkelbach}}, \bibinfo {author} {\bibfnamefont {M.~S.}\ \bibnamefont
  {Hybertsen}}, \ and\ \bibinfo {author} {\bibfnamefont {D.~R.}\ \bibnamefont
  {Reichman}},\ }\href {\doibase 10.1063/1.4794427} {\bibfield  {journal}
  {\bibinfo  {journal} {J. Chem. Phys.}\ }\textbf {\bibinfo {volume} {138}},\
  \bibinfo {pages} {114103} (\bibinfo {year} {2013}{\natexlab{a}})}\BibitemShut
  {NoStop}%
\bibitem [{\citenamefont {Zimmerman}\ \emph {et~al.}(2011)\citenamefont
  {Zimmerman}, \citenamefont {Bell}, \citenamefont {Casanova},\ and\
  \citenamefont {Head-Gordon}}]{zimmerman_mechanism_2011}%
  \BibitemOpen
  \bibfield  {author} {\bibinfo {author} {\bibfnamefont {P.~M.}\ \bibnamefont
  {Zimmerman}}, \bibinfo {author} {\bibfnamefont {F.}~\bibnamefont {Bell}},
  \bibinfo {author} {\bibfnamefont {D.}~\bibnamefont {Casanova}}, \ and\
  \bibinfo {author} {\bibfnamefont {M.}~\bibnamefont {Head-Gordon}},\ }\href
  {\doibase 10.1021/ja208431r} {\bibfield  {journal} {\bibinfo  {journal} {J.
  Am. Chem. Soc.}\ }\textbf {\bibinfo {volume} {133}},\ \bibinfo {pages}
  {19944} (\bibinfo {year} {2011})}\BibitemShut {NoStop}%
\bibitem [{\citenamefont {Zimmerman}\ \emph {et~al.}(2010)\citenamefont
  {Zimmerman}, \citenamefont {Zhang},\ and\ \citenamefont
  {Musgrave}}]{zimmerman_singlet_2010}%
  \BibitemOpen
  \bibfield  {author} {\bibinfo {author} {\bibfnamefont {P.~M.}\ \bibnamefont
  {Zimmerman}}, \bibinfo {author} {\bibfnamefont {Z.}~\bibnamefont {Zhang}}, \
  and\ \bibinfo {author} {\bibfnamefont {C.~B.}\ \bibnamefont {Musgrave}},\
  }\href {\doibase 10.1038/nchem.694} {\bibfield  {journal} {\bibinfo
  {journal} {Nat. Chem.}\ }\textbf {\bibinfo {volume} {2}},\ \bibinfo {pages}
  {648} (\bibinfo {year} {2010})}\BibitemShut {NoStop}%
\bibitem [{\citenamefont {Zeng}\ \emph {et~al.}(2014)\citenamefont {Zeng},
  \citenamefont {Hoffmann},\ and\ \citenamefont
  {Ananth}}]{zeng_low-lying_2014}%
  \BibitemOpen
  \bibfield  {author} {\bibinfo {author} {\bibfnamefont {T.}~\bibnamefont
  {Zeng}}, \bibinfo {author} {\bibfnamefont {R.}~\bibnamefont {Hoffmann}}, \
  and\ \bibinfo {author} {\bibfnamefont {N.}~\bibnamefont {Ananth}},\ }\href
  {\doibase 10.1021/ja500887a} {\bibfield  {journal} {\bibinfo  {journal} {J.
  Am. Chem. Soc.}\ }\textbf {\bibinfo {volume} {136}},\ \bibinfo {pages} {5755}
  (\bibinfo {year} {2014})}\BibitemShut {NoStop}%
\bibitem [{\citenamefont {Congreve}\ \emph {et~al.}(2013)\citenamefont
  {Congreve}, \citenamefont {Lee}, \citenamefont {Thompson}, \citenamefont
  {Hontz}, \citenamefont {Yost}, \citenamefont {Reusswig}, \citenamefont
  {Bahlke}, \citenamefont {Reineke}, \citenamefont {Voorhis},\ and\
  \citenamefont {Baldo}}]{congreve_external_2013}%
  \BibitemOpen
  \bibfield  {author} {\bibinfo {author} {\bibfnamefont {D.~N.}\ \bibnamefont
  {Congreve}}, \bibinfo {author} {\bibfnamefont {J.}~\bibnamefont {Lee}},
  \bibinfo {author} {\bibfnamefont {N.~J.}\ \bibnamefont {Thompson}}, \bibinfo
  {author} {\bibfnamefont {E.}~\bibnamefont {Hontz}}, \bibinfo {author}
  {\bibfnamefont {S.~R.}\ \bibnamefont {Yost}}, \bibinfo {author}
  {\bibfnamefont {P.~D.}\ \bibnamefont {Reusswig}}, \bibinfo {author}
  {\bibfnamefont {M.~E.}\ \bibnamefont {Bahlke}}, \bibinfo {author}
  {\bibfnamefont {S.}~\bibnamefont {Reineke}}, \bibinfo {author} {\bibfnamefont
  {T.~V.}\ \bibnamefont {Voorhis}}, \ and\ \bibinfo {author} {\bibfnamefont
  {M.~A.}\ \bibnamefont {Baldo}},\ }\href {\doibase 10.1126/science.1232994}
  {\bibfield  {journal} {\bibinfo  {journal} {Science}\ }\textbf {\bibinfo
  {volume} {340}},\ \bibinfo {pages} {334} (\bibinfo {year}
  {2013})}\BibitemShut {NoStop}%
\bibitem [{\citenamefont {Bakulin}\ \emph {et~al.}(2016)\citenamefont
  {Bakulin}, \citenamefont {Morgan}, \citenamefont {Kehoe}, \citenamefont
  {Wilson}, \citenamefont {Chin}, \citenamefont {Zigmantas}, \citenamefont
  {Egorova},\ and\ \citenamefont {Rao}}]{bakulin_real-time_2016}%
  \BibitemOpen
  \bibfield  {author} {\bibinfo {author} {\bibfnamefont {A.~A.}\ \bibnamefont
  {Bakulin}}, \bibinfo {author} {\bibfnamefont {S.~E.}\ \bibnamefont {Morgan}},
  \bibinfo {author} {\bibfnamefont {T.~B.}\ \bibnamefont {Kehoe}}, \bibinfo
  {author} {\bibfnamefont {M.~W.~B.}\ \bibnamefont {Wilson}}, \bibinfo {author}
  {\bibfnamefont {A.~W.}\ \bibnamefont {Chin}}, \bibinfo {author}
  {\bibfnamefont {D.}~\bibnamefont {Zigmantas}}, \bibinfo {author}
  {\bibfnamefont {D.}~\bibnamefont {Egorova}}, \ and\ \bibinfo {author}
  {\bibfnamefont {A.}~\bibnamefont {Rao}},\ }\href {\doibase
  10.1038/nchem.2371} {\bibfield  {journal} {\bibinfo  {journal} {Nat. Chem.}\
  }\textbf {\bibinfo {volume} {8}},\ \bibinfo {pages} {16} (\bibinfo {year}
  {2016})}\BibitemShut {NoStop}%
\bibitem [{\citenamefont {Berkelbach}\ \emph
  {et~al.}(2013{\natexlab{b}})\citenamefont {Berkelbach}, \citenamefont
  {Hybertsen},\ and\ \citenamefont {Reichman}}]{berkelbach_microscopic_2013-1}%
  \BibitemOpen
  \bibfield  {author} {\bibinfo {author} {\bibfnamefont {T.~C.}\ \bibnamefont
  {Berkelbach}}, \bibinfo {author} {\bibfnamefont {M.~S.}\ \bibnamefont
  {Hybertsen}}, \ and\ \bibinfo {author} {\bibfnamefont {D.~R.}\ \bibnamefont
  {Reichman}},\ }\href {\doibase 10.1063/1.4794425} {\bibfield  {journal}
  {\bibinfo  {journal} {J. Chem. Phys.}\ }\textbf {\bibinfo {volume} {138}},\
  \bibinfo {pages} {114102} (\bibinfo {year} {2013}{\natexlab{b}})}\BibitemShut
  {NoStop}%
\bibitem [{\citenamefont {Berkelbach}\ \emph {et~al.}(2014)\citenamefont
  {Berkelbach}, \citenamefont {Hybertsen},\ and\ \citenamefont
  {Reichman}}]{berkelbach_microscopic_2014}%
  \BibitemOpen
  \bibfield  {author} {\bibinfo {author} {\bibfnamefont {T.~C.}\ \bibnamefont
  {Berkelbach}}, \bibinfo {author} {\bibfnamefont {M.~S.}\ \bibnamefont
  {Hybertsen}}, \ and\ \bibinfo {author} {\bibfnamefont {D.~R.}\ \bibnamefont
  {Reichman}},\ }\href {\doibase 10.1063/1.4892793} {\bibfield  {journal}
  {\bibinfo  {journal} {J. Chem. Phys.}\ }\textbf {\bibinfo {volume} {141}},\
  \bibinfo {pages} {074705} (\bibinfo {year} {2014})}\BibitemShut {NoStop}%
\bibitem [{\citenamefont {Chan}\ \emph {et~al.}(2013)\citenamefont {Chan},
  \citenamefont {Berkelbach}, \citenamefont {Provorse}, \citenamefont
  {Monahan}, \citenamefont {Tritsch}, \citenamefont {Hybertsen}, \citenamefont
  {Reichman}, \citenamefont {Gao},\ and\ \citenamefont
  {Zhu}}]{chan_quantum_2013}%
  \BibitemOpen
  \bibfield  {author} {\bibinfo {author} {\bibfnamefont {W.-L.}\ \bibnamefont
  {Chan}}, \bibinfo {author} {\bibfnamefont {T.~C.}\ \bibnamefont
  {Berkelbach}}, \bibinfo {author} {\bibfnamefont {M.~R.}\ \bibnamefont
  {Provorse}}, \bibinfo {author} {\bibfnamefont {N.~R.}\ \bibnamefont
  {Monahan}}, \bibinfo {author} {\bibfnamefont {J.~R.}\ \bibnamefont
  {Tritsch}}, \bibinfo {author} {\bibfnamefont {M.~S.}\ \bibnamefont
  {Hybertsen}}, \bibinfo {author} {\bibfnamefont {D.~R.}\ \bibnamefont
  {Reichman}}, \bibinfo {author} {\bibfnamefont {J.}~\bibnamefont {Gao}}, \
  and\ \bibinfo {author} {\bibfnamefont {X.-Y.}\ \bibnamefont {Zhu}},\ }\href
  {\doibase 10.1021/ar300286s} {\bibfield  {journal} {\bibinfo  {journal} {Acc.
  Chem. Res.}\ }\textbf {\bibinfo {volume} {46}},\ \bibinfo {pages} {1321}
  (\bibinfo {year} {2013})}\BibitemShut {NoStop}%
\bibitem [{\citenamefont {Sebastian}\ \emph {et~al.}(1981)\citenamefont
  {Sebastian}, \citenamefont {Weiser},\ and\ \citenamefont
  {B\"assler}}]{sebastian_charge_1981}%
  \BibitemOpen
  \bibfield  {author} {\bibinfo {author} {\bibfnamefont {L.}~\bibnamefont
  {Sebastian}}, \bibinfo {author} {\bibfnamefont {G.}~\bibnamefont {Weiser}}, \
  and\ \bibinfo {author} {\bibfnamefont {H.}~\bibnamefont {B\"assler}},\ }\href
  {\doibase 10.1016/0301-0104(81)85055-0} {\bibfield  {journal} {\bibinfo
  {journal} {Chem. Phys.}\ }\textbf {\bibinfo {volume} {61}},\ \bibinfo {pages}
  {125} (\bibinfo {year} {1981})}\BibitemShut {NoStop}%
\bibitem [{\citenamefont {Sebastian}\ \emph {et~al.}(1983)\citenamefont
  {Sebastian}, \citenamefont {Weiser}, \citenamefont {Peter},\ and\
  \citenamefont {B\"assler}}]{sebastian_charge-transfer_1983}%
  \BibitemOpen
  \bibfield  {author} {\bibinfo {author} {\bibfnamefont {L.}~\bibnamefont
  {Sebastian}}, \bibinfo {author} {\bibfnamefont {G.}~\bibnamefont {Weiser}},
  \bibinfo {author} {\bibfnamefont {G.}~\bibnamefont {Peter}}, \ and\ \bibinfo
  {author} {\bibfnamefont {H.}~\bibnamefont {B\"assler}},\ }\href {\doibase
  10.1016/0301-0104(83)85012-5} {\bibfield  {journal} {\bibinfo  {journal}
  {Chem. Phys.}\ }\textbf {\bibinfo {volume} {75}},\ \bibinfo {pages} {103}
  (\bibinfo {year} {1983})}\BibitemShut {NoStop}%
\bibitem [{\citenamefont {Dreuw}\ \emph {et~al.}(2003)\citenamefont {Dreuw},
  \citenamefont {Weisman},\ and\ \citenamefont
  {Head-Gordon}}]{dreuw_long-range_2003}%
  \BibitemOpen
  \bibfield  {author} {\bibinfo {author} {\bibfnamefont {A.}~\bibnamefont
  {Dreuw}}, \bibinfo {author} {\bibfnamefont {J.~L.}\ \bibnamefont {Weisman}},
  \ and\ \bibinfo {author} {\bibfnamefont {M.}~\bibnamefont {Head-Gordon}},\
  }\href {\doibase 10.1063/1.1590951} {\bibfield  {journal} {\bibinfo
  {journal} {J. Chem. Phys.}\ }\textbf {\bibinfo {volume} {119}},\ \bibinfo
  {pages} {2943} (\bibinfo {year} {2003})}\BibitemShut {NoStop}%
\bibitem [{\citenamefont {Tawada}\ \emph {et~al.}(2004)\citenamefont {Tawada},
  \citenamefont {Tsuneda}, \citenamefont {Yanagisawa}, \citenamefont {Yanai},\
  and\ \citenamefont {Hirao}}]{tawada_long-range-corrected_2004}%
  \BibitemOpen
  \bibfield  {author} {\bibinfo {author} {\bibfnamefont {Y.}~\bibnamefont
  {Tawada}}, \bibinfo {author} {\bibfnamefont {T.}~\bibnamefont {Tsuneda}},
  \bibinfo {author} {\bibfnamefont {S.}~\bibnamefont {Yanagisawa}}, \bibinfo
  {author} {\bibfnamefont {T.}~\bibnamefont {Yanai}}, \ and\ \bibinfo {author}
  {\bibfnamefont {K.}~\bibnamefont {Hirao}},\ }\href {\doibase
  10.1063/1.1688752} {\bibfield  {journal} {\bibinfo  {journal} {J. Chem.
  Phys.}\ }\textbf {\bibinfo {volume} {120}},\ \bibinfo {pages} {8425}
  (\bibinfo {year} {2004})}\BibitemShut {NoStop}%
\bibitem [{\citenamefont {Sharifzadeh}\ \emph {et~al.}(2012)\citenamefont
  {Sharifzadeh}, \citenamefont {Biller}, \citenamefont {Kronik},\ and\
  \citenamefont {Neaton}}]{sharifzadeh_quasiparticle_2012}%
  \BibitemOpen
  \bibfield  {author} {\bibinfo {author} {\bibfnamefont {S.}~\bibnamefont
  {Sharifzadeh}}, \bibinfo {author} {\bibfnamefont {A.}~\bibnamefont {Biller}},
  \bibinfo {author} {\bibfnamefont {L.}~\bibnamefont {Kronik}}, \ and\ \bibinfo
  {author} {\bibfnamefont {J.~B.}\ \bibnamefont {Neaton}},\ }\href {\doibase
  10.1103/PhysRevB.85.125307} {\bibfield  {journal} {\bibinfo  {journal} {Phys.
  Rev. B}\ }\textbf {\bibinfo {volume} {85}},\ \bibinfo {pages} {125307}
  (\bibinfo {year} {2012})}\BibitemShut {NoStop}%
\bibitem [{\citenamefont {Dederichs}\ \emph {et~al.}(1984)\citenamefont
  {Dederichs}, \citenamefont {Bl\"ugel}, \citenamefont {Zeller},\ and\
  \citenamefont {Akai}}]{dederichs_ground_1984}%
  \BibitemOpen
  \bibfield  {author} {\bibinfo {author} {\bibfnamefont {P.~H.}\ \bibnamefont
  {Dederichs}}, \bibinfo {author} {\bibfnamefont {S.}~\bibnamefont {Bl\"ugel}},
  \bibinfo {author} {\bibfnamefont {R.}~\bibnamefont {Zeller}}, \ and\ \bibinfo
  {author} {\bibfnamefont {H.}~\bibnamefont {Akai}},\ }\href {\doibase
  10.1103/PhysRevLett.53.2512} {\bibfield  {journal} {\bibinfo  {journal}
  {Phys. Rev. Lett.}\ }\textbf {\bibinfo {volume} {53}},\ \bibinfo {pages}
  {2512} (\bibinfo {year} {1984})}\BibitemShut {NoStop}%
\bibitem [{\citenamefont {Wu}\ and\ \citenamefont
  {Van~Voorhis}(2005)}]{wu_direct_2005}%
  \BibitemOpen
  \bibfield  {author} {\bibinfo {author} {\bibfnamefont {Q.}~\bibnamefont
  {Wu}}\ and\ \bibinfo {author} {\bibfnamefont {T.}~\bibnamefont
  {Van~Voorhis}},\ }\href {\doibase 10.1103/PhysRevA.72.024502} {\bibfield
  {journal} {\bibinfo  {journal} {Phys. Rev. A}\ }\textbf {\bibinfo {volume}
  {72}},\ \bibinfo {pages} {024502} (\bibinfo {year} {2005})}\BibitemShut
  {NoStop}%
\bibitem [{\citenamefont {Wu}\ and\ \citenamefont
  {Van~Voorhis}(2006)}]{wu_constrained_2006}%
  \BibitemOpen
  \bibfield  {author} {\bibinfo {author} {\bibfnamefont {Q.}~\bibnamefont
  {Wu}}\ and\ \bibinfo {author} {\bibfnamefont {T.}~\bibnamefont
  {Van~Voorhis}},\ }\href {\doibase 10.1021/ct0503163} {\bibfield  {journal}
  {\bibinfo  {journal} {J. Chem. Theory}\ }\textbf {\bibinfo {volume} {2}},\
  \bibinfo {pages} {765} (\bibinfo {year} {2006})}\BibitemShut {NoStop}%
\bibitem [{\citenamefont {Kaduk}\ \emph {et~al.}(2012)\citenamefont {Kaduk},
  \citenamefont {Kowalczyk},\ and\ \citenamefont
  {Van~Voorhis}}]{kaduk_constrained_2012}%
  \BibitemOpen
  \bibfield  {author} {\bibinfo {author} {\bibfnamefont {B.}~\bibnamefont
  {Kaduk}}, \bibinfo {author} {\bibfnamefont {T.}~\bibnamefont {Kowalczyk}}, \
  and\ \bibinfo {author} {\bibfnamefont {T.}~\bibnamefont {Van~Voorhis}},\
  }\href {\doibase 10.1021/cr200148b} {\bibfield  {journal} {\bibinfo
  {journal} {Chem. Rev.}\ }\textbf {\bibinfo {volume} {112}},\ \bibinfo {pages}
  {321} (\bibinfo {year} {2012})}\BibitemShut {NoStop}%
\bibitem [{\citenamefont {Skylaris}\ \emph {et~al.}(2005)\citenamefont
  {Skylaris}, \citenamefont {Haynes}, \citenamefont {Mostofi},\ and\
  \citenamefont {Payne}}]{skylaris_introducing_2005}%
  \BibitemOpen
  \bibfield  {author} {\bibinfo {author} {\bibfnamefont {C.-K.}\ \bibnamefont
  {Skylaris}}, \bibinfo {author} {\bibfnamefont {P.~D.}\ \bibnamefont
  {Haynes}}, \bibinfo {author} {\bibfnamefont {A.~A.}\ \bibnamefont {Mostofi}},
  \ and\ \bibinfo {author} {\bibfnamefont {M.~C.}\ \bibnamefont {Payne}},\
  }\href {\doibase 10.1063/1.1839852} {\bibfield  {journal} {\bibinfo
  {journal} {J. Chem. Phys.}\ }\textbf {\bibinfo {volume} {122}},\ \bibinfo
  {pages} {084119} (\bibinfo {year} {2005})}\BibitemShut {NoStop}%
\bibitem [{\citenamefont {\u{R}ez\'a\u{c}}\ and\ \citenamefont {de~la
  Lande}(2015)}]{rezac_robust_2015}%
  \BibitemOpen
  \bibfield  {author} {\bibinfo {author} {\bibfnamefont {J.}~\bibnamefont
  {\u{R}ez\'a\u{c}}}\ and\ \bibinfo {author} {\bibfnamefont {A.}~\bibnamefont
  {de~la Lande}},\ }\href {\doibase 10.1021/ct501115m} {\bibfield  {journal}
  {\bibinfo  {journal} {J. Chem. Theory}\ }\textbf {\bibinfo {volume} {11}},\
  \bibinfo {pages} {528} (\bibinfo {year} {2015})}\BibitemShut {NoStop}%
\bibitem [{\citenamefont {Vaissier}\ \emph {et~al.}(2015)\citenamefont
  {Vaissier}, \citenamefont {Frost}, \citenamefont {Barnes},\ and\
  \citenamefont {Nelson}}]{vaissier_influence_2015}%
  \BibitemOpen
  \bibfield  {author} {\bibinfo {author} {\bibfnamefont {V.}~\bibnamefont
  {Vaissier}}, \bibinfo {author} {\bibfnamefont {J.~M.}\ \bibnamefont {Frost}},
  \bibinfo {author} {\bibfnamefont {P.~R.~F.}\ \bibnamefont {Barnes}}, \ and\
  \bibinfo {author} {\bibfnamefont {J.}~\bibnamefont {Nelson}},\ }\href
  {\doibase 10.1021/acs.jpcc.5b09739} {\bibfield  {journal} {\bibinfo
  {journal} {J. Phys. Chem. C}\ }\textbf {\bibinfo {volume} {119}},\ \bibinfo
  {pages} {24337} (\bibinfo {year} {2015})}\BibitemShut {NoStop}%
\bibitem [{\citenamefont {Zheng}\ \emph {et~al.}(2012)\citenamefont {Zheng},
  \citenamefont {Phillips}, \citenamefont {Geva},\ and\ \citenamefont
  {Dunietz}}]{zheng_ab_2012}%
  \BibitemOpen
  \bibfield  {author} {\bibinfo {author} {\bibfnamefont {S.}~\bibnamefont
  {Zheng}}, \bibinfo {author} {\bibfnamefont {H.}~\bibnamefont {Phillips}},
  \bibinfo {author} {\bibfnamefont {E.}~\bibnamefont {Geva}}, \ and\ \bibinfo
  {author} {\bibfnamefont {B.~D.}\ \bibnamefont {Dunietz}},\ }\href {\doibase
  10.1021/ja301442v} {\bibfield  {journal} {\bibinfo  {journal} {J. Am. Chem.
  Soc.}\ }\textbf {\bibinfo {volume} {134}},\ \bibinfo {pages} {6944} (\bibinfo
  {year} {2012})}\BibitemShut {NoStop}%
\bibitem [{\citenamefont {Zheng}\ \emph {et~al.}(2013)\citenamefont {Zheng},
  \citenamefont {Geva},\ and\ \citenamefont {Dunietz}}]{zheng_solvated_2013}%
  \BibitemOpen
  \bibfield  {author} {\bibinfo {author} {\bibfnamefont {S.}~\bibnamefont
  {Zheng}}, \bibinfo {author} {\bibfnamefont {E.}~\bibnamefont {Geva}}, \ and\
  \bibinfo {author} {\bibfnamefont {B.~D.}\ \bibnamefont {Dunietz}},\ }\href
  {\doibase 10.1021/ct300700q} {\bibfield  {journal} {\bibinfo  {journal} {J.
  Chem. Theory}\ }\textbf {\bibinfo {volume} {9}},\ \bibinfo {pages} {1125}
  (\bibinfo {year} {2013})}\BibitemShut {NoStop}%
\bibitem [{\citenamefont {Kubas}\ \emph {et~al.}(2014)\citenamefont {Kubas},
  \citenamefont {Hoffmann}, \citenamefont {Heck}, \citenamefont {Oberhofer},
  \citenamefont {Elstner},\ and\ \citenamefont
  {Blumberger}}]{kubas_electronic_2014}%
  \BibitemOpen
  \bibfield  {author} {\bibinfo {author} {\bibfnamefont {A.}~\bibnamefont
  {Kubas}}, \bibinfo {author} {\bibfnamefont {F.}~\bibnamefont {Hoffmann}},
  \bibinfo {author} {\bibfnamefont {A.}~\bibnamefont {Heck}}, \bibinfo {author}
  {\bibfnamefont {H.}~\bibnamefont {Oberhofer}}, \bibinfo {author}
  {\bibfnamefont {M.}~\bibnamefont {Elstner}}, \ and\ \bibinfo {author}
  {\bibfnamefont {J.}~\bibnamefont {Blumberger}},\ }\href {\doibase
  10.1063/1.4867077} {\bibfield  {journal} {\bibinfo  {journal} {J. Chem.
  Phys.}\ }\textbf {\bibinfo {volume} {140}},\ \bibinfo {pages} {104105}
  (\bibinfo {year} {2014})}\BibitemShut {NoStop}%
\bibitem [{\citenamefont {Kubas}\ \emph {et~al.}(2015)\citenamefont {Kubas},
  \citenamefont {Gajdos}, \citenamefont {Heck}, \citenamefont {Oberhofer},
  \citenamefont {Elstner},\ and\ \citenamefont
  {Blumberger}}]{kubas_electronic_2015}%
  \BibitemOpen
  \bibfield  {author} {\bibinfo {author} {\bibfnamefont {A.}~\bibnamefont
  {Kubas}}, \bibinfo {author} {\bibfnamefont {F.}~\bibnamefont {Gajdos}},
  \bibinfo {author} {\bibfnamefont {A.}~\bibnamefont {Heck}}, \bibinfo {author}
  {\bibfnamefont {H.}~\bibnamefont {Oberhofer}}, \bibinfo {author}
  {\bibfnamefont {M.}~\bibnamefont {Elstner}}, \ and\ \bibinfo {author}
  {\bibfnamefont {J.}~\bibnamefont {Blumberger}},\ }\href {\doibase
  10.1039/C4CP04749D} {\bibfield  {journal} {\bibinfo  {journal} {Phys. Chem.
  Chem. Phys.}\ }\textbf {\bibinfo {volume} {17}},\ \bibinfo {pages} {14342}
  (\bibinfo {year} {2015})}\BibitemShut {NoStop}%
\bibitem [{\citenamefont {Si}\ \emph {et~al.}(2012)\citenamefont {Si},
  \citenamefont {Liang},\ and\ \citenamefont {Zhao}}]{si_theoretical_2012}%
  \BibitemOpen
  \bibfield  {author} {\bibinfo {author} {\bibfnamefont {Y.}~\bibnamefont
  {Si}}, \bibinfo {author} {\bibfnamefont {W.}~\bibnamefont {Liang}}, \ and\
  \bibinfo {author} {\bibfnamefont {Y.}~\bibnamefont {Zhao}},\ }\href {\doibase
  10.1021/jp303705d} {\bibfield  {journal} {\bibinfo  {journal} {J. Phys. Chem.
  C}\ }\textbf {\bibinfo {volume} {116}},\ \bibinfo {pages} {12499} (\bibinfo
  {year} {2012})}\BibitemShut {NoStop}%
\bibitem [{\citenamefont {Aikawa}\ \emph {et~al.}(2015)\citenamefont {Aikawa},
  \citenamefont {Sumita}, \citenamefont {Shimodo},\ and\ \citenamefont
  {Morihashi}}]{aikawa_theoretical_2015}%
  \BibitemOpen
  \bibfield  {author} {\bibinfo {author} {\bibfnamefont {K.}~\bibnamefont
  {Aikawa}}, \bibinfo {author} {\bibfnamefont {M.}~\bibnamefont {Sumita}},
  \bibinfo {author} {\bibfnamefont {Y.}~\bibnamefont {Shimodo}}, \ and\
  \bibinfo {author} {\bibfnamefont {K.}~\bibnamefont {Morihashi}},\ }\href
  {\doibase 10.1039/C5CP03235K} {\bibfield  {journal} {\bibinfo  {journal}
  {Phys. Chem. Chem. Phys.}\ }\textbf {\bibinfo {volume} {17}},\ \bibinfo
  {pages} {20923} (\bibinfo {year} {2015})}\BibitemShut {NoStop}%
\bibitem [{\citenamefont {Eisenmayer}\ \emph {et~al.}(2013)\citenamefont
  {Eisenmayer}, \citenamefont {Lasave}, \citenamefont {Monti}, \citenamefont
  {de~Groot},\ and\ \citenamefont {Buda}}]{eisenmayer_proton_2013}%
  \BibitemOpen
  \bibfield  {author} {\bibinfo {author} {\bibfnamefont {T.~J.}\ \bibnamefont
  {Eisenmayer}}, \bibinfo {author} {\bibfnamefont {J.~A.}\ \bibnamefont
  {Lasave}}, \bibinfo {author} {\bibfnamefont {A.}~\bibnamefont {Monti}},
  \bibinfo {author} {\bibfnamefont {H.~J.~M.}\ \bibnamefont {de~Groot}}, \ and\
  \bibinfo {author} {\bibfnamefont {F.}~\bibnamefont {Buda}},\ }\href {\doibase
  10.1021/jp401195t} {\bibfield  {journal} {\bibinfo  {journal} {J. Phys. Chem.
  B}\ }\textbf {\bibinfo {volume} {117}},\ \bibinfo {pages} {11162} (\bibinfo
  {year} {2013})}\BibitemShut {NoStop}%
\bibitem [{\citenamefont {Yu}\ \emph {et~al.}(2012)\citenamefont {Yu},
  \citenamefont {Huang}, \citenamefont {Shapter},\ and\ \citenamefont
  {Abell}}]{yu_electrochemical_2012}%
  \BibitemOpen
  \bibfield  {author} {\bibinfo {author} {\bibfnamefont {J.}~\bibnamefont
  {Yu}}, \bibinfo {author} {\bibfnamefont {D.~M.}\ \bibnamefont {Huang}},
  \bibinfo {author} {\bibfnamefont {J.~G.}\ \bibnamefont {Shapter}}, \ and\
  \bibinfo {author} {\bibfnamefont {A.~D.}\ \bibnamefont {Abell}},\ }\href
  {\doibase 10.1021/jp3082563} {\bibfield  {journal} {\bibinfo  {journal} {J.
  Phys. Chem. C}\ }\textbf {\bibinfo {volume} {116}},\ \bibinfo {pages} {26608}
  (\bibinfo {year} {2012})}\BibitemShut {NoStop}%
\bibitem [{\citenamefont {Siefermann}\ \emph {et~al.}(2014)\citenamefont
  {Siefermann}, \citenamefont {Pemmaraju}, \citenamefont {Neppl}, \citenamefont
  {Shavorskiy}, \citenamefont {Cordones}, \citenamefont {Vura-Weis},
  \citenamefont {Slaughter}, \citenamefont {Sturm}, \citenamefont {Weise},
  \citenamefont {Bluhm}, \citenamefont {Strader}, \citenamefont {Cho},
  \citenamefont {Lin}, \citenamefont {Bacellar}, \citenamefont {Khurmi},
  \citenamefont {Guo}, \citenamefont {Coslovich}, \citenamefont {Robinson},
  \citenamefont {Kaindl}, \citenamefont {Schoenlein}, \citenamefont {Belkacem},
  \citenamefont {Neumark}, \citenamefont {Leone}, \citenamefont {Nordlund},
  \citenamefont {Ogasawara}, \citenamefont {Krupin}, \citenamefont {Turner},
  \citenamefont {Schlotter}, \citenamefont {Holmes}, \citenamefont
  {Messerschmidt}, \citenamefont {Minitti}, \citenamefont {Gul}, \citenamefont
  {Zhang}, \citenamefont {Huse}, \citenamefont {Prendergast},\ and\
  \citenamefont {Gessner}}]{siefermann_atomic-scale_2014}%
  \BibitemOpen
  \bibfield  {author} {\bibinfo {author} {\bibfnamefont {K.~R.}\ \bibnamefont
  {Siefermann}}, \bibinfo {author} {\bibfnamefont {C.~D.}\ \bibnamefont
  {Pemmaraju}}, \bibinfo {author} {\bibfnamefont {S.}~\bibnamefont {Neppl}},
  \bibinfo {author} {\bibfnamefont {A.}~\bibnamefont {Shavorskiy}}, \bibinfo
  {author} {\bibfnamefont {A.~A.}\ \bibnamefont {Cordones}}, \bibinfo {author}
  {\bibfnamefont {J.}~\bibnamefont {Vura-Weis}}, \bibinfo {author}
  {\bibfnamefont {D.~S.}\ \bibnamefont {Slaughter}}, \bibinfo {author}
  {\bibfnamefont {F.~P.}\ \bibnamefont {Sturm}}, \bibinfo {author}
  {\bibfnamefont {F.}~\bibnamefont {Weise}}, \bibinfo {author} {\bibfnamefont
  {H.}~\bibnamefont {Bluhm}}, \bibinfo {author} {\bibfnamefont {M.~L.}\
  \bibnamefont {Strader}}, \bibinfo {author} {\bibfnamefont {H.}~\bibnamefont
  {Cho}}, \bibinfo {author} {\bibfnamefont {M.-F.}\ \bibnamefont {Lin}},
  \bibinfo {author} {\bibfnamefont {C.}~\bibnamefont {Bacellar}}, \bibinfo
  {author} {\bibfnamefont {C.}~\bibnamefont {Khurmi}}, \bibinfo {author}
  {\bibfnamefont {J.}~\bibnamefont {Guo}}, \bibinfo {author} {\bibfnamefont
  {G.}~\bibnamefont {Coslovich}}, \bibinfo {author} {\bibfnamefont {J.~S.}\
  \bibnamefont {Robinson}}, \bibinfo {author} {\bibfnamefont {R.~A.}\
  \bibnamefont {Kaindl}}, \bibinfo {author} {\bibfnamefont {R.~W.}\
  \bibnamefont {Schoenlein}}, \bibinfo {author} {\bibfnamefont
  {A.}~\bibnamefont {Belkacem}}, \bibinfo {author} {\bibfnamefont {D.~M.}\
  \bibnamefont {Neumark}}, \bibinfo {author} {\bibfnamefont {S.~R.}\
  \bibnamefont {Leone}}, \bibinfo {author} {\bibfnamefont {D.}~\bibnamefont
  {Nordlund}}, \bibinfo {author} {\bibfnamefont {H.}~\bibnamefont {Ogasawara}},
  \bibinfo {author} {\bibfnamefont {O.}~\bibnamefont {Krupin}}, \bibinfo
  {author} {\bibfnamefont {J.~J.}\ \bibnamefont {Turner}}, \bibinfo {author}
  {\bibfnamefont {W.~F.}\ \bibnamefont {Schlotter}}, \bibinfo {author}
  {\bibfnamefont {M.~R.}\ \bibnamefont {Holmes}}, \bibinfo {author}
  {\bibfnamefont {M.}~\bibnamefont {Messerschmidt}}, \bibinfo {author}
  {\bibfnamefont {M.~P.}\ \bibnamefont {Minitti}}, \bibinfo {author}
  {\bibfnamefont {S.}~\bibnamefont {Gul}}, \bibinfo {author} {\bibfnamefont
  {J.~Z.}\ \bibnamefont {Zhang}}, \bibinfo {author} {\bibfnamefont
  {N.}~\bibnamefont {Huse}}, \bibinfo {author} {\bibfnamefont {D.}~\bibnamefont
  {Prendergast}}, \ and\ \bibinfo {author} {\bibfnamefont {O.}~\bibnamefont
  {Gessner}},\ }\href {\doibase 10.1021/jz501264x} {\bibfield  {journal}
  {\bibinfo  {journal} {J. Phys. Chem. Letters}\ }\textbf {\bibinfo {volume}
  {5}},\ \bibinfo {pages} {2753} (\bibinfo {year} {2014})}\BibitemShut
  {NoStop}%
\bibitem [{\citenamefont {\u{R}ez\'a\u{c}}\ \emph {et~al.}(2012)\citenamefont
  {\u{R}ez\'a\u{c}}, \citenamefont {L\'evy}, \citenamefont {Demachy},\ and\
  \citenamefont {de~la Lande}}]{rezac_robust_2012}%
  \BibitemOpen
  \bibfield  {author} {\bibinfo {author} {\bibfnamefont {J.}~\bibnamefont
  {\u{R}ez\'a\u{c}}}, \bibinfo {author} {\bibfnamefont {B.}~\bibnamefont
  {L\'evy}}, \bibinfo {author} {\bibfnamefont {I.}~\bibnamefont {Demachy}}, \
  and\ \bibinfo {author} {\bibfnamefont {A.}~\bibnamefont {de~la Lande}},\
  }\href {\doibase 10.1021/ct200570u} {\bibfield  {journal} {\bibinfo
  {journal} {J. Chem. Theory}\ }\textbf {\bibinfo {volume} {8}},\ \bibinfo
  {pages} {418} (\bibinfo {year} {2012})}\BibitemShut {NoStop}%
\bibitem [{\citenamefont {Oberhofer}\ and\ \citenamefont
  {Blumberger}(2009)}]{oberhofer_charge_2009}%
  \BibitemOpen
  \bibfield  {author} {\bibinfo {author} {\bibfnamefont {H.}~\bibnamefont
  {Oberhofer}}\ and\ \bibinfo {author} {\bibfnamefont {J.}~\bibnamefont
  {Blumberger}},\ }\href {\doibase 10.1063/1.3190169} {\bibfield  {journal}
  {\bibinfo  {journal} {J. Chem. Phys.}\ }\textbf {\bibinfo {volume} {131}},\
  \bibinfo {pages} {064101} (\bibinfo {year} {2009})}\BibitemShut {NoStop}%
\bibitem [{\citenamefont {Sun}\ and\ \citenamefont
  {Dalton}(2008)}]{sun_introduction_2008}%
  \BibitemOpen
  \bibfield  {author} {\bibinfo {author} {\bibfnamefont {S.-S.}\ \bibnamefont
  {Sun}}\ and\ \bibinfo {author} {\bibfnamefont {L.~R.}\ \bibnamefont
  {Dalton}},\ }\href
  {https://www.crcpress.com/Introduction-to-Organic-Electronic-and-Optoelectronic-Materials-and-Devices/Sun-Dalton/9780849392849}
  {\emph {\bibinfo {title} {Introduction to {Organic} {Electronic} and
  {Optoelectronic} {Materials} and {Devices}}}}\ (\bibinfo  {publisher} {CRC
  Press: New York},\ \bibinfo {year} {2008})\BibitemShut {NoStop}%
\bibitem [{\citenamefont {Forrest}(2004)}]{forrest_path_2004}%
  \BibitemOpen
  \bibfield  {author} {\bibinfo {author} {\bibfnamefont {S.~R.}\ \bibnamefont
  {Forrest}},\ }\href {\doibase 10.1038/nature02498} {\bibfield  {journal}
  {\bibinfo  {journal} {Nature}\ }\textbf {\bibinfo {volume} {428}},\ \bibinfo
  {pages} {911} (\bibinfo {year} {2004})}\BibitemShut {NoStop}%
\bibitem [{\citenamefont {Zhao}\ \emph {et~al.}(2013)\citenamefont {Zhao},
  \citenamefont {Guo},\ and\ \citenamefont {Liu}}]{zhao_25th_2013}%
  \BibitemOpen
  \bibfield  {author} {\bibinfo {author} {\bibfnamefont {Y.}~\bibnamefont
  {Zhao}}, \bibinfo {author} {\bibfnamefont {Y.}~\bibnamefont {Guo}}, \ and\
  \bibinfo {author} {\bibfnamefont {Y.}~\bibnamefont {Liu}},\ }\href {\doibase
  10.1002/adma.201302315} {\bibfield  {journal} {\bibinfo  {journal} {Adv.
  Mater.}\ }\textbf {\bibinfo {volume} {25}},\ \bibinfo {pages} {5372}
  (\bibinfo {year} {2013})}\BibitemShut {NoStop}%
\bibitem [{\citenamefont {Naber}\ \emph {et~al.}(2007)\citenamefont {Naber},
  \citenamefont {Faez},\ and\ \citenamefont {Wiel}}]{naber_organic_2007}%
  \BibitemOpen
  \bibfield  {author} {\bibinfo {author} {\bibfnamefont {W.~J.~M.}\
  \bibnamefont {Naber}}, \bibinfo {author} {\bibfnamefont {S.}~\bibnamefont
  {Faez}}, \ and\ \bibinfo {author} {\bibfnamefont {W.~G. v.~d.}\ \bibnamefont
  {Wiel}},\ }\href {\doibase 10.1088/0022-3727/40/12/R01} {\bibfield  {journal}
  {\bibinfo  {journal} {J. Phys. D}\ }\textbf {\bibinfo {volume} {40}},\
  \bibinfo {pages} {R205} (\bibinfo {year} {2007})}\BibitemShut {NoStop}%
\bibitem [{\citenamefont {Baroni}\ \emph {et~al.}(2001)\citenamefont {Baroni},
  \citenamefont {de~Gironcoli}, \citenamefont {Dal~Corso},\ and\ \citenamefont
  {Giannozzi}}]{baroni_phonons_2001}%
  \BibitemOpen
  \bibfield  {author} {\bibinfo {author} {\bibfnamefont {S.}~\bibnamefont
  {Baroni}}, \bibinfo {author} {\bibfnamefont {S.}~\bibnamefont
  {de~Gironcoli}}, \bibinfo {author} {\bibfnamefont {A.}~\bibnamefont
  {Dal~Corso}}, \ and\ \bibinfo {author} {\bibfnamefont {P.}~\bibnamefont
  {Giannozzi}},\ }\href {\doibase 10.1103/RevModPhys.73.515} {\bibfield
  {journal} {\bibinfo  {journal} {Rev. Mod. Phys.}\ }\textbf {\bibinfo {volume}
  {73}},\ \bibinfo {pages} {515} (\bibinfo {year} {2001})}\BibitemShut
  {NoStop}%
\bibitem [{\citenamefont {Coto}\ \emph {et~al.}(2014)\citenamefont {Coto},
  \citenamefont {Sharifzadeh}, \citenamefont {Neaton},\ and\ \citenamefont
  {Thoss}}]{coto_low-lying_2014}%
  \BibitemOpen
  \bibfield  {author} {\bibinfo {author} {\bibfnamefont {P.~B.}\ \bibnamefont
  {Coto}}, \bibinfo {author} {\bibfnamefont {S.}~\bibnamefont {Sharifzadeh}},
  \bibinfo {author} {\bibfnamefont {J.~B.}\ \bibnamefont {Neaton}}, \ and\
  \bibinfo {author} {\bibfnamefont {M.}~\bibnamefont {Thoss}},\ }\href
  {\doibase 10.1021/ct500510k} {\bibfield  {journal} {\bibinfo  {journal} {J.
  Chem. Theory}\ }\textbf {\bibinfo {volume} {11}},\ \bibinfo {pages} {147}
  (\bibinfo {year} {2014})}\BibitemShut {NoStop}%
\bibitem [{\citenamefont {Skylaris}\ \emph {et~al.}(2002)\citenamefont
  {Skylaris}, \citenamefont {Mostofi}, \citenamefont {Haynes}, \citenamefont
  {Di\'eguez},\ and\ \citenamefont {Payne}}]{skylaris_nonorthogonal_2002}%
  \BibitemOpen
  \bibfield  {author} {\bibinfo {author} {\bibfnamefont {C.-K.}\ \bibnamefont
  {Skylaris}}, \bibinfo {author} {\bibfnamefont {A.~A.}\ \bibnamefont
  {Mostofi}}, \bibinfo {author} {\bibfnamefont {P.~D.}\ \bibnamefont {Haynes}},
  \bibinfo {author} {\bibfnamefont {O.}~\bibnamefont {Di\'eguez}}, \ and\
  \bibinfo {author} {\bibfnamefont {M.~C.}\ \bibnamefont {Payne}},\ }\href
  {\doibase 10.1103/PhysRevB.66.035119} {\bibfield  {journal} {\bibinfo
  {journal} {Phys. Rev. B}\ }\textbf {\bibinfo {volume} {66}},\ \bibinfo
  {pages} {035119} (\bibinfo {year} {2002})}\BibitemShut {NoStop}%
\bibitem [{\citenamefont {Kohn}(2008)}]{kohn_nearsightedness_2008}%
  \BibitemOpen
  \bibfield  {author} {\bibinfo {author} {\bibfnamefont {W.}~\bibnamefont
  {Kohn}},\ }\enquote {\bibinfo {title} {Nearsightedness of electronic
  matter},}\ in\ \href@noop {} {\emph {\bibinfo {booktitle} {Proceedings of the
  Conference in Honor of C N Yang's 85th Birthday}}}\ (\bibinfo  {publisher}
  {World Scientific},\ \bibinfo {year} {2008})\ Chap.~\bibinfo {chapter} {18},
  pp.\ \bibinfo {pages} {217--217}\BibitemShut {NoStop}%
\bibitem [{\citenamefont {Skylaris}\ and\ \citenamefont
  {Haynes}(2007)}]{skylaris_achieving_2007}%
  \BibitemOpen
  \bibfield  {author} {\bibinfo {author} {\bibfnamefont {C.-K.}\ \bibnamefont
  {Skylaris}}\ and\ \bibinfo {author} {\bibfnamefont {P.~D.}\ \bibnamefont
  {Haynes}},\ }\href {\doibase 10.1063/1.2796168} {\bibfield  {journal}
  {\bibinfo  {journal} {J. Chem. Phys.}\ }\textbf {\bibinfo {volume} {127}},\
  \bibinfo {pages} {164712} (\bibinfo {year} {2007})}\BibitemShut {NoStop}%
\bibitem [{\citenamefont {O'Regan}\ \emph {et~al.}(2010)\citenamefont
  {O'Regan}, \citenamefont {Hine}, \citenamefont {Payne},\ and\ \citenamefont
  {Mostofi}}]{oregan_projector_2010}%
  \BibitemOpen
  \bibfield  {author} {\bibinfo {author} {\bibfnamefont {D.~D.}\ \bibnamefont
  {O'Regan}}, \bibinfo {author} {\bibfnamefont {N.~D.~M.}\ \bibnamefont
  {Hine}}, \bibinfo {author} {\bibfnamefont {M.~C.}\ \bibnamefont {Payne}}, \
  and\ \bibinfo {author} {\bibfnamefont {A.~A.}\ \bibnamefont {Mostofi}},\
  }\href {\doibase 10.1103/PhysRevB.82.081102} {\bibfield  {journal} {\bibinfo
  {journal} {Phys. Rev. B}\ }\textbf {\bibinfo {volume} {82}},\ \bibinfo
  {pages} {081102} (\bibinfo {year} {2010})}\BibitemShut {NoStop}%
\bibitem [{\citenamefont {Artacho}\ and\ \citenamefont {Mil\'{a}ns~del
  Bosch}(1991)}]{artacho_nonorthogonal_1991}%
  \BibitemOpen
  \bibfield  {author} {\bibinfo {author} {\bibfnamefont {E.}~\bibnamefont
  {Artacho}}\ and\ \bibinfo {author} {\bibfnamefont {L.}~\bibnamefont
  {Mil\'{a}ns~del Bosch}},\ }\href@noop {} {\bibfield  {journal} {\bibinfo
  {journal} {Phys. Rev. A}\ }\textbf {\bibinfo {volume} {43}},\ \bibinfo
  {pages} {5770} (\bibinfo {year} {1991})}\BibitemShut {NoStop}%
\bibitem [{\citenamefont {O'Regan}\ \emph {et~al.}(2012)\citenamefont
  {O'Regan}, \citenamefont {Hine}, \citenamefont {Payne},\ and\ \citenamefont
  {Mostofi}}]{oregan_linear-scaling_2012}%
  \BibitemOpen
  \bibfield  {author} {\bibinfo {author} {\bibfnamefont {D.~D.}\ \bibnamefont
  {O'Regan}}, \bibinfo {author} {\bibfnamefont {N.~D.~M.}\ \bibnamefont
  {Hine}}, \bibinfo {author} {\bibfnamefont {M.~C.}\ \bibnamefont {Payne}}, \
  and\ \bibinfo {author} {\bibfnamefont {A.~A.}\ \bibnamefont {Mostofi}},\
  }\href {\doibase 10.1103/PhysRevB.85.085107} {\bibfield  {journal} {\bibinfo
  {journal} {Phys. Rev. B}\ }\textbf {\bibinfo {volume} {85}},\ \bibinfo
  {pages} {085107} (\bibinfo {year} {2012})}\BibitemShut {NoStop}%
\bibitem [{\citenamefont {Ozaki}(2001)}]{ozaki_efficient_2001}%
  \BibitemOpen
  \bibfield  {author} {\bibinfo {author} {\bibfnamefont {T.}~\bibnamefont
  {Ozaki}},\ }\href {\doibase 10.1103/PhysRevB.64.195110} {\bibfield  {journal}
  {\bibinfo  {journal} {Phys. Rev. B}\ }\textbf {\bibinfo {volume} {64}},\
  \bibinfo {pages} {195110} (\bibinfo {year} {2001})}\BibitemShut {NoStop}%
\bibitem [{\citenamefont {O'Regan}\ \emph {et~al.}(2011)\citenamefont
  {O'Regan}, \citenamefont {Payne},\ and\ \citenamefont
  {Mostofi}}]{oregan_subspace_2011}%
  \BibitemOpen
  \bibfield  {author} {\bibinfo {author} {\bibfnamefont {D.~D.}\ \bibnamefont
  {O'Regan}}, \bibinfo {author} {\bibfnamefont {M.~C.}\ \bibnamefont {Payne}},
  \ and\ \bibinfo {author} {\bibfnamefont {A.~A.}\ \bibnamefont {Mostofi}},\
  }\href {\doibase 10.1103/PhysRevB.83.245124} {\bibfield  {journal} {\bibinfo
  {journal} {Phys. Rev. B}\ }\textbf {\bibinfo {volume} {83}},\ \bibinfo
  {pages} {245124} (\bibinfo {year} {2011})}\BibitemShut {NoStop}%
\bibitem [{\citenamefont {Ratcliff}\ \emph
  {et~al.}(2015{\natexlab{a}})\citenamefont {Ratcliff}, \citenamefont
  {Grisanti}, \citenamefont {Genovese}, \citenamefont {Deutsch}, \citenamefont
  {Neumann}, \citenamefont {Danilov}, \citenamefont {Wenzel}, \citenamefont
  {Beljonne},\ and\ \citenamefont {Cornil}}]{ratcliff_toward_2015}%
  \BibitemOpen
  \bibfield  {author} {\bibinfo {author} {\bibfnamefont {L.~E.}\ \bibnamefont
  {Ratcliff}}, \bibinfo {author} {\bibfnamefont {L.}~\bibnamefont {Grisanti}},
  \bibinfo {author} {\bibfnamefont {L.}~\bibnamefont {Genovese}}, \bibinfo
  {author} {\bibfnamefont {T.}~\bibnamefont {Deutsch}}, \bibinfo {author}
  {\bibfnamefont {T.}~\bibnamefont {Neumann}}, \bibinfo {author} {\bibfnamefont
  {D.}~\bibnamefont {Danilov}}, \bibinfo {author} {\bibfnamefont
  {W.}~\bibnamefont {Wenzel}}, \bibinfo {author} {\bibfnamefont
  {D.}~\bibnamefont {Beljonne}}, \ and\ \bibinfo {author} {\bibfnamefont
  {J.}~\bibnamefont {Cornil}},\ }\href {\doibase 10.1021/acs.jctc.5b00057}
  {\bibfield  {journal} {\bibinfo  {journal} {J. Chem. Theory}\ }\textbf
  {\bibinfo {volume} {11}},\ \bibinfo {pages} {2077} (\bibinfo {year}
  {2015}{\natexlab{a}})}\BibitemShut {NoStop}%
\bibitem [{\citenamefont {Ratcliff}\ \emph
  {et~al.}(2015{\natexlab{b}})\citenamefont {Ratcliff}, \citenamefont
  {Genovese}, \citenamefont {Mohr},\ and\ \citenamefont
  {Deutsch}}]{ratcliff_fragment_2015}%
  \BibitemOpen
  \bibfield  {author} {\bibinfo {author} {\bibfnamefont {L.~E.}\ \bibnamefont
  {Ratcliff}}, \bibinfo {author} {\bibfnamefont {L.}~\bibnamefont {Genovese}},
  \bibinfo {author} {\bibfnamefont {S.}~\bibnamefont {Mohr}}, \ and\ \bibinfo
  {author} {\bibfnamefont {T.}~\bibnamefont {Deutsch}},\ }\href {\doibase
  10.1063/1.4922378} {\bibfield  {journal} {\bibinfo  {journal} {J. Chem.
  Phys.}\ }\textbf {\bibinfo {volume} {142}},\ \bibinfo {pages} {234105}
  (\bibinfo {year} {2015}{\natexlab{b}})}\BibitemShut {NoStop}%
\bibitem [{\citenamefont {Ruiz-Serrano}\ \emph {et~al.}(2012)\citenamefont
  {Ruiz-Serrano}, \citenamefont {Hine},\ and\ \citenamefont
  {Skylaris}}]{ruiz-serrano_pulay_2012}%
  \BibitemOpen
  \bibfield  {author} {\bibinfo {author} {\bibfnamefont {A.}~\bibnamefont
  {Ruiz-Serrano}}, \bibinfo {author} {\bibfnamefont {N.~D.~M.}\ \bibnamefont
  {Hine}}, \ and\ \bibinfo {author} {\bibfnamefont {C.-K.}\ \bibnamefont
  {Skylaris}},\ }\href {\doibase 10.1063/1.4728026} {\bibfield  {journal}
  {\bibinfo  {journal} {J. Chem. Phys.}\ }\textbf {\bibinfo {volume} {136}},\
  \bibinfo {pages} {234101} (\bibinfo {year} {2012})}\BibitemShut {NoStop}%
\bibitem [{\citenamefont {Clark}\ \emph {et~al.}(2005)\citenamefont {Clark},
  \citenamefont {Segall}, \citenamefont {Pickard}, \citenamefont {Hasnip},
  \citenamefont {Probert}, \citenamefont {Refson},\ and\ \citenamefont
  {Payne}}]{clark_first_2009}%
  \BibitemOpen
  \bibfield  {author} {\bibinfo {author} {\bibfnamefont {S.~J.}\ \bibnamefont
  {Clark}}, \bibinfo {author} {\bibfnamefont {M.~D.}\ \bibnamefont {Segall}},
  \bibinfo {author} {\bibfnamefont {C.~J.}\ \bibnamefont {Pickard}}, \bibinfo
  {author} {\bibfnamefont {P.~J.}\ \bibnamefont {Hasnip}}, \bibinfo {author}
  {\bibfnamefont {M.~I.~J.}\ \bibnamefont {Probert}}, \bibinfo {author}
  {\bibfnamefont {K.}~\bibnamefont {Refson}}, \ and\ \bibinfo {author}
  {\bibfnamefont {M.~C.}\ \bibnamefont {Payne}},\ }\href {\doibase
  10.1524/zkri.220.5.567.65075} {\bibfield  {journal} {\bibinfo  {journal} {Z.
  Kristallogr.}\ }\textbf {\bibinfo {volume} {220}},\ \bibinfo {pages} {567}
  (\bibinfo {year} {2005})}\BibitemShut {NoStop}%
\bibitem [{\citenamefont {Campbell}\ \emph {et~al.}(1962)\citenamefont
  {Campbell}, \citenamefont {Robertson},\ and\ \citenamefont
  {Trotter}}]{campbell_crystal_1962}%
  \BibitemOpen
  \bibfield  {author} {\bibinfo {author} {\bibfnamefont {R.~B.}\ \bibnamefont
  {Campbell}}, \bibinfo {author} {\bibfnamefont {J.~M.}\ \bibnamefont
  {Robertson}}, \ and\ \bibinfo {author} {\bibfnamefont {J.}~\bibnamefont
  {Trotter}},\ }\href {\doibase 10.1107/S0365110X62000699} {\bibfield
  {journal} {\bibinfo  {journal} {Acta Crystallogr.}\ }\textbf {\bibinfo
  {volume} {15}},\ \bibinfo {pages} {289} (\bibinfo {year} {1962})}\BibitemShut
  {NoStop}%
\bibitem [{\citenamefont {Hine}\ \emph {et~al.}(2011)\citenamefont {Hine},
  \citenamefont {Dziedzic}, \citenamefont {Haynes},\ and\ \citenamefont
  {Skylaris}}]{hine_electrostatic_2011}%
  \BibitemOpen
  \bibfield  {author} {\bibinfo {author} {\bibfnamefont {N.~D.~M.}\
  \bibnamefont {Hine}}, \bibinfo {author} {\bibfnamefont {J.}~\bibnamefont
  {Dziedzic}}, \bibinfo {author} {\bibfnamefont {P.~D.}\ \bibnamefont
  {Haynes}}, \ and\ \bibinfo {author} {\bibfnamefont {C.-K.}\ \bibnamefont
  {Skylaris}},\ }\href {\doibase 10.1063/1.3662863} {\bibfield  {journal}
  {\bibinfo  {journal} {J. Chem. Phys.}\ }\textbf {\bibinfo {volume} {135}},\
  \bibinfo {pages} {204103} (\bibinfo {year} {2011})}\BibitemShut {NoStop}%
\bibitem [{\citenamefont {Ewald}(1921)}]{ewald_berechnung_1921}%
  \BibitemOpen
  \bibfield  {author} {\bibinfo {author} {\bibfnamefont {P.~P.}\ \bibnamefont
  {Ewald}},\ }\href {\doibase 10.1002/andp.19213690304} {\bibfield  {journal}
  {\bibinfo  {journal} {Ann. Phys.}\ }\textbf {\bibinfo {volume} {369}},\
  \bibinfo {pages} {253} (\bibinfo {year} {1921})}\BibitemShut {NoStop}%
\bibitem [{\citenamefont {Makov}\ and\ \citenamefont
  {Payne}(1995)}]{makov_periodic_1995}%
  \BibitemOpen
  \bibfield  {author} {\bibinfo {author} {\bibfnamefont {G.}~\bibnamefont
  {Makov}}\ and\ \bibinfo {author} {\bibfnamefont {M.~C.}\ \bibnamefont
  {Payne}},\ }\href {\doibase 10.1103/PhysRevB.51.4014} {\bibfield  {journal}
  {\bibinfo  {journal} {Phys. Rev. B}\ }\textbf {\bibinfo {volume} {51}},\
  \bibinfo {pages} {4014} (\bibinfo {year} {1995})}\BibitemShut {NoStop}%
\bibitem [{\citenamefont {Leslie}\ and\ \citenamefont
  {Gillan}(1985)}]{leslie_energy_1985}%
  \BibitemOpen
  \bibfield  {author} {\bibinfo {author} {\bibfnamefont {M.}~\bibnamefont
  {Leslie}}\ and\ \bibinfo {author} {\bibfnamefont {N.~J.}\ \bibnamefont
  {Gillan}},\ }\href {\doibase 10.1088/0022-3719/18/5/005} {\bibfield
  {journal} {\bibinfo  {journal} {J. Phys. C: Solid State Phys.}\ }\textbf
  {\bibinfo {volume} {18}},\ \bibinfo {pages} {973} (\bibinfo {year}
  {1985})}\BibitemShut {NoStop}%
\bibitem [{\citenamefont {McKenna}\ and\ \citenamefont
  {Blumberger}(2012)}]{mckenna_crossover_2012}%
  \BibitemOpen
  \bibfield  {author} {\bibinfo {author} {\bibfnamefont {K.~P.}\ \bibnamefont
  {McKenna}}\ and\ \bibinfo {author} {\bibfnamefont {J.}~\bibnamefont
  {Blumberger}},\ }\href {\doibase 10.1103/PhysRevB.86.245110} {\bibfield
  {journal} {\bibinfo  {journal} {Phys. Rev. B}\ }\textbf {\bibinfo {volume}
  {86}},\ \bibinfo {pages} {245110} (\bibinfo {year} {2012})}\BibitemShut
  {NoStop}%
\bibitem [{\citenamefont {Blumberger}\ and\ \citenamefont
  {McKenna}(2013)}]{blumberger_constrained_2013}%
  \BibitemOpen
  \bibfield  {author} {\bibinfo {author} {\bibfnamefont {J.}~\bibnamefont
  {Blumberger}}\ and\ \bibinfo {author} {\bibfnamefont {K.~P.}\ \bibnamefont
  {McKenna}},\ }\href {\doibase 10.1039/C2CP42537H} {\bibfield  {journal}
  {\bibinfo  {journal} {Phys. Chem. Chem. Phys.}\ }\textbf {\bibinfo {volume}
  {15}},\ \bibinfo {pages} {2184} (\bibinfo {year} {2013})}\BibitemShut
  {NoStop}%
\bibitem [{\citenamefont {Murphy}\ and\ \citenamefont
  {Hine}(2013)}]{murphy_anisotropic_2013}%
  \BibitemOpen
  \bibfield  {author} {\bibinfo {author} {\bibfnamefont {S.~T.}\ \bibnamefont
  {Murphy}}\ and\ \bibinfo {author} {\bibfnamefont {N.~D.~M.}\ \bibnamefont
  {Hine}},\ }\href {\doibase 10.1103/PhysRevB.87.094111} {\bibfield  {journal}
  {\bibinfo  {journal} {Phys. Rev. B}\ }\textbf {\bibinfo {volume} {87}},\
  \bibinfo {pages} {094111} (\bibinfo {year} {2013})}\BibitemShut {NoStop}%
\bibitem [{\citenamefont {Kantorovich}(1999)}]{kantorovich_elimination_1999}%
  \BibitemOpen
  \bibfield  {author} {\bibinfo {author} {\bibfnamefont {L.~N.}\ \bibnamefont
  {Kantorovich}},\ }\href {\doibase 10.1103/PhysRevB.60.15476} {\bibfield
  {journal} {\bibinfo  {journal} {Phys. Rev. B}\ }\textbf {\bibinfo {volume}
  {60}},\ \bibinfo {pages} {15476} (\bibinfo {year} {1999})}\BibitemShut
  {NoStop}%
\bibitem [{\citenamefont {Fischerauer}(1997)}]{fischerauer_comments_1997}%
  \BibitemOpen
  \bibfield  {author} {\bibinfo {author} {\bibfnamefont {G.}~\bibnamefont
  {Fischerauer}},\ }\href {\doibase 10.1109/58.656617} {\bibfield  {journal}
  {\bibinfo  {journal} {IEEE Trans. Ultrason, Ferroelect., Freq. Control}\
  }\textbf {\bibinfo {volume} {44}},\ \bibinfo {pages} {1179} (\bibinfo {year}
  {1997})}\BibitemShut {NoStop}%
\bibitem [{\citenamefont {Lever}\ \emph {et~al.}(2013)\citenamefont {Lever},
  \citenamefont {Cole}, \citenamefont {Hine}, \citenamefont {Haynes},\ and\
  \citenamefont {Payne}}]{lever_electrostatic_2013}%
  \BibitemOpen
  \bibfield  {author} {\bibinfo {author} {\bibfnamefont {G.}~\bibnamefont
  {Lever}}, \bibinfo {author} {\bibfnamefont {D.~J.}\ \bibnamefont {Cole}},
  \bibinfo {author} {\bibfnamefont {N.~D.~M.}\ \bibnamefont {Hine}}, \bibinfo
  {author} {\bibfnamefont {P.~D.}\ \bibnamefont {Haynes}}, \ and\ \bibinfo
  {author} {\bibfnamefont {M.~C.}\ \bibnamefont {Payne}},\ }\href {\doibase
  10.1088/0953-8984/25/15/152101} {\bibfield  {journal} {\bibinfo  {journal}
  {J. Phys. Condens. Matter}\ }\textbf {\bibinfo {volume} {25}},\ \bibinfo
  {pages} {152101} (\bibinfo {year} {2013})}\BibitemShut {NoStop}%
\bibitem [{\citenamefont {Tiago}\ \emph {et~al.}(2003)\citenamefont {Tiago},
  \citenamefont {Northrup},\ and\ \citenamefont {Louie}}]{tiago_ab_2003}%
  \BibitemOpen
  \bibfield  {author} {\bibinfo {author} {\bibfnamefont {M.~L.}\ \bibnamefont
  {Tiago}}, \bibinfo {author} {\bibfnamefont {J.~E.}\ \bibnamefont {Northrup}},
  \ and\ \bibinfo {author} {\bibfnamefont {S.~G.}\ \bibnamefont {Louie}},\
  }\href {\doibase 10.1103/PhysRevB.67.115212} {\bibfield  {journal} {\bibinfo
  {journal} {Phys. Rev. B}\ }\textbf {\bibinfo {volume} {67}},\ \bibinfo
  {pages} {115212} (\bibinfo {year} {2003})}\BibitemShut {NoStop}%
\bibitem [{\citenamefont {Cudazzo}\ \emph {et~al.}(2012)\citenamefont
  {Cudazzo}, \citenamefont {Gatti},\ and\ \citenamefont
  {Rubio}}]{cudazzo_excitons_2012}%
  \BibitemOpen
  \bibfield  {author} {\bibinfo {author} {\bibfnamefont {P.}~\bibnamefont
  {Cudazzo}}, \bibinfo {author} {\bibfnamefont {M.}~\bibnamefont {Gatti}}, \
  and\ \bibinfo {author} {\bibfnamefont {A.}~\bibnamefont {Rubio}},\ }\href
  {\doibase 10.1103/PhysRevB.86.195307} {\bibfield  {journal} {\bibinfo
  {journal} {Phys. Rev. B}\ }\textbf {\bibinfo {volume} {86}},\ \bibinfo
  {pages} {195307} (\bibinfo {year} {2012})}\BibitemShut {NoStop}%
\bibitem [{\citenamefont {Sharifzadeh}\ \emph {et~al.}(2013)\citenamefont
  {Sharifzadeh}, \citenamefont {Darancet}, \citenamefont {Kronik},\ and\
  \citenamefont {Neaton}}]{sharifzadeh_low-energy_2013}%
  \BibitemOpen
  \bibfield  {author} {\bibinfo {author} {\bibfnamefont {S.}~\bibnamefont
  {Sharifzadeh}}, \bibinfo {author} {\bibfnamefont {P.}~\bibnamefont
  {Darancet}}, \bibinfo {author} {\bibfnamefont {L.}~\bibnamefont {Kronik}}, \
  and\ \bibinfo {author} {\bibfnamefont {J.~B.}\ \bibnamefont {Neaton}},\
  }\href {\doibase 10.1021/jz401069f} {\bibfield  {journal} {\bibinfo
  {journal} {J. Phys. Chem. Letters}\ }\textbf {\bibinfo {volume} {4}},\
  \bibinfo {pages} {2197} (\bibinfo {year} {2013})}\BibitemShut {NoStop}%
\end{thebibliography}%

\end{document}